\documentclass{article}

\usepackage{PRIMEarxiv}

\usepackage[utf8]{inputenc} 
\usepackage[T1]{fontenc}    
\usepackage{url}            
\usepackage{booktabs}       
\usepackage{amsfonts}       
\usepackage{nicefrac}       
\usepackage{microtype}      
\usepackage{lipsum}
\usepackage{fancyhdr}       
\usepackage{graphicx}       
\graphicspath{{media/}}     
\usepackage{subcaption}
\usepackage{amsmath}
\usepackage{bm}
\usepackage{listings}
\usepackage{makecell}
\usepackage{float}
\usepackage{anyfontsize}
\DeclareUnicodeCharacter{2113}{l}
\title{KAN/MultKAN with Physics-Informed Spline fitting (KAN-PISF) for ordinary/partial differential equation discovery of nonlinear dynamic systems
}

\author{
  Ashish Pal\\
  Department of Civil and Environmental Engineering \\
  Rice University \\
  Houston, Texas, USA, 77005\\
  \texttt{ashish.pal@rice.edu} \\
   \AND
  Satish Nagarajaiah \\
  Department of Civil and Environmental Engineering \\
  Department of Mechanical Engineering \\
  Rice University \\
  Houston, Texas, USA, 77005\\
  \texttt{satish.nagarajaiah@rice.edu} \\
}

\begin{document}
\maketitle

\begin{abstract}
Machine learning for scientific discovery is increasingly becoming popular because of its ability to extract and recognize the nonlinear characteristics from the data. The black-box nature of deep learning methods poses difficulties in interpreting the identified model. There is a dire need to interpret the machine learning models to develop a physical understanding of dynamic systems.  An interpretable form of neural network called Kolmogorov-Arnold networks (KAN) or Multiplicative KAN (MultKAN) offers critical features that help recognize the nonlinearities in the governing ordinary/partial differential equations (ODE/PDE) of various dynamic systems and find their equation structures. In this study, an equation discovery framework is proposed that includes i) sequentially regularized derivatives for denoising (SRDD) algorithm to denoise the measure data to obtain accurate derivatives, ii) KAN to identify the equation structure and suggest relevant nonlinear functions that are used to create a small overcomplete library of functions, and iii) physics-informed spline fitting (PISF) algorithm to filter the excess functions from the library and converge to the correct equation. The framework was tested on the forced Duffing oscillator, Van der Pol oscillator (stiff ODE), Burger's equation, and Bouc-Wen model (coupled ODE). The proposed method converged to the true equation for the first three systems. It provided an approximate model for the Bouc-Wen model that could acceptably capture the hysteresis response. Using KAN maintains low complexity, which helps the user interpret the results throughout the process and avoid the black-box-type nature of machine learning methods. 
\end{abstract}

\keywords{Kolmogorov-Arnold Networks \and Physics-informed spline fitting \and SRDD \and Nonlinear system identification \and Equation discovery}

\section{INTRODUCTION}
Nonlinear dynamics is commonly found in engineering systems and is often modeled using ODEs/PDEs \cite{verhulst2012nonlinear,perko2013differential,hirsch2013differential}. The ODE/PDE discovery is traditionally done via first principles that primarily rely on expert knowledge and experience. The demand for experiments might significantly increase based on the analyst interpretation and inferences. These physical demands and limitations might hinder in fully understanding the unknown underlying nonlinear dynamics of the systems. With the advent of advanced data acquisition systems, computational power, and machine learning algorithms, a promising alternative for nonlinear system identification is now possible \cite{lai2019semi,bhowmick2023physics,rudy2017data,brunton2022data}.\\
Various data-driven methods have been developed over the years for nonlinear system identification that involves Koopman operator \cite{williams2015data}, subspace methods \cite{marchesiello2008time}, dynamic mode decomposition \cite{proctor2016dynamic}, and machine learning methods \cite{ayala2020nonlinear,yu2019system,silvestrini2022deep,wu2019deep,de2016randomized,li2020fourier}. Although these methods capture the nonlinear characteristics of dynamic systems, physical interpretation might not always be possible and extrapolation of the dynamic response may be unreliable beyond the training data space. In recent years, interpretable nonlinear system identification methods have been proposed that directly target to identify the ODEs/PDEs guiding the nonlinear behavior of the physical systems \cite{brunton2016discovering,schaeffer2017learning,nayek2021spike,pal2024physics,chen2022symbolic}. These methods include, Kalman-filter-based methods \cite{pal2024sparsity,zhang2017structural}, physics-informed machine learning \cite{long2018pde,long2019pde,both2021deepmod,xu2021robust,raviprakash2022hybrid,stephany2022pde,raissi2018hidden,raissi2019physics,sharma2023review,zhang2022parsimony}, physics-informed spline fitting \cite{bhowmick2023physics,bhowmick2023data,pal2024physics}, symbolic regression \cite{bomarito2021development,reinbold2021robust,sun2019data,sun2024data}, and regression-based sparse-system identification \cite{cortiella2021sparse,lin2021nonlinear,champion2019data,rudy2017data}.\\
Broadly, these methods can be classified into two categories: dictionary-based and dictionary-free. Each of these classes of methods has its own advantages and disadvantages. In dictionary-based methods, a library of nonlinear functions of system response and its partial derivatives is created. To ensure that the library contains the relevant functions, many possible nonlinear functions must be included to form an overcomplete representation. Having a large library makes converging to true functions difficult due to false functions correlating well with them and replacing them \cite{pal2024physics}. Also, it is difficult to ensure that true functions are indeed part of the library, which requires insight into the system's behavior and choosing the functions that would help capture it. In dictionary-free methods, the nonlinear functions and the structure of the equation are automatically selected with no input from the user. However, these methods suffer from high computational expense for large problems and a significant risk of overfitting for a bigger network. There is low reliability in the equation obtained from such methods, and the overfitting is also difficult to control. In conclusion, one class suffers from a lack of automatic nonlinear function selection, while the other suffers from unreliability due to overfitting. Another problem common to both and indeed to any system identification method is the noise in the measured data. This can particularly damage equation discovery methods as the derivatives are calculated numerically. If the derivatives contain significant errors, then it is not possible to derive the correct equation. All of these issues are addressed in this study to achieve a robust equation-discovery framework.\\
In this study, we propose a framework in which the equation structure and relevant nonlinear functions are found using a new type of neural network named Kolmogorov-Arnold networks (KAN) \cite{liu2024kan}. In contrast to the common Multi-layer Perceptron (MLP), which has fixed activations and learnable scalar weights, KAN has learnable activation functions that form the basis of interpretable networks. The inputs to KAN go through a series of learnable activation functions, which can be added and multiplied to generate the output. The equation structure can be obtained by unfolding the network based on the addition and multiplication of the activation functions. Once the equation structure is determined, the next task is to find the correct nonlinear functions. Another property of KAN is that it does not fix the nonlinear functions as in symbolic regression; rather, it learns the nonlinear functions and suggests a list of best-matching functions for them. These suggestions of nonlinear functions and equation structure from KAN are then used to form a small dictionary of functions that can be used to converge to the correct unknown ODE/PDE. The challenging task of handling the noise in the measured data is done with the help of two algorithms, SRDD and PISF \cite{pal2024physics}. The SRDD algorithm is used to initially denoise the data to get good-quality derivatives from the noisy measurement that are fit enough to discover the network structure and relevant nonlinear functions. In the machine learning framework, automatic differentiation is a convenient way to find the derivatives. However, due to the noise in the signal, the exact differentiation of the output leads to erroneous derivatives, which cannot be used for equation discovery. The derivatives obtained from SRDD are found to be much superior compared to automatic differentiation. PISF is used in the later stage to eliminate the low-contributing functions from the dictionary sequentially. This algorithm simultaneously fits the measurement data and the governing equation based on the available dictionary. As the dictionary becomes smaller, simultaneous fitting reshapes the fitting to the measured signal while changing the derivatives accordingly, gradually converging to the true shape even in the presence of high noise.\\
The structure of the rest of the paper is as follows: (i) background on KAN and its helpful modification Multiplicative KAN (MultKAN), (ii) develop the equation discovery framework using the forced Duffing oscillator ODE, (iii) test it on Van der Pol oscillator, Burger's equation, and Bouc-Wen model, and (iv) Conclusions and future scope.

\section{METHODOLOGY}
\subsection{KAN: Kolmogorov-Arnold Networks}
\begin{figure}
\centering
  \includegraphics[width=1\linewidth]{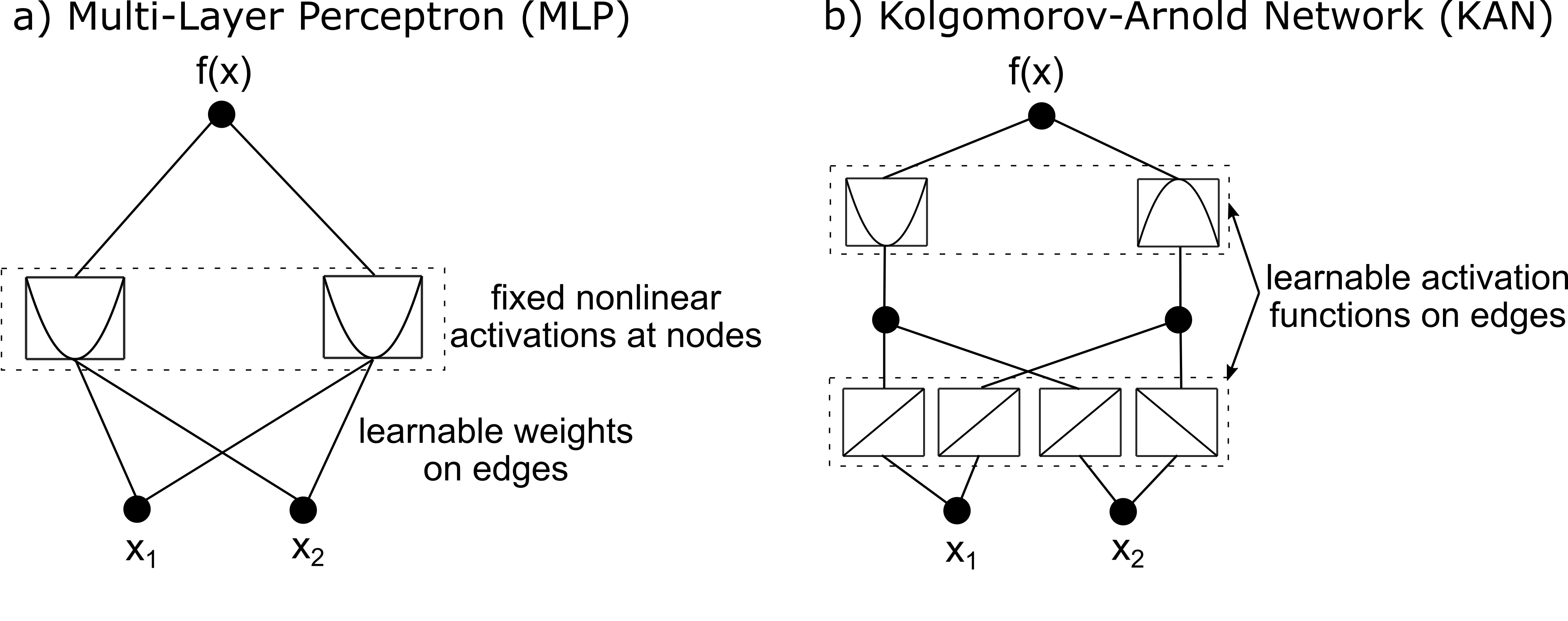}
\caption{Comparison between the MLP and KAN architecture}
\label{fig:MLP_KAN}
\end{figure}

Recently, KAN \cite{liu2024kan} has been introduced as an alternative to Multi-Layer Perceptron (MLP). In MLP, the weights connecting nodes from one layer to another are scalar and learnable, and the nodes have a fixed nonlinear activation. In KAN, there are no scalar weights; instead, the weights connecting nodes from one layer to another are learnable univariate functions parameterized as a spline. The advantage of KAN is that it offers interpretable networks compared to MLP. This is illustrated using an example case in Figure \ref{fig:MLP_KAN} showing the fundamental difference between the two types of networks. In Figure \ref{fig:MLP_KAN}a, a two-layer MLP with (2,2,1) nodes is shown whose output can be written as the following:

\begin{equation}
    f(x) = \sum_{i=1}^2w_{2,1i}\Phi\Bigg(\sum_{j=1}^2w_{1,ij}x_j+b_i\Bigg)
\end{equation}
where, $\Phi$ is a fixed nonlinear activation functions, and $w_{l,ij}$ is the weight in the $l^\mathrm{th}$ layer connecting the $i^\mathrm{th}$ output node with the $j^\mathrm{th}$ input node. Now assume we are training an MLP where the inputs are $x_1$, $x_2$, and the output is $4x_1x_2$. For the MLP illustrated in Figure \ref{fig:MLP_KAN}, if the activation function is set to $x^2$, then the following weights, $w_{1,1j}=[1,1]$, $w_{1,2j}=[1,-1]$, and $w_{2,1j}=[1,-1]$ will lead to $f(x)=(x_1+x_2)^2-(x_1-x_2)^2=4x_1x_2$, capturing the exact relation between inputs and outputs in just two layers. However, knowing which nonlinear activation functions should be used for an unknown system is generally nearly impossible. The conventional nonlinear activation functions used in MLPs do not even include quadratic nonlinear activations. To capture the input-output relation for the above example using the conventional nonlinear activation, the MLPs must be made deeper, which results in the loss of interpretability of the network.\\
Now, looking at how the KAN works, Figure \ref{fig:MLP_KAN}b shows a two-layer KAN with (2,2,1) nodes whose output can be written as the following:
\begin{equation}
    f(x) = \sum_{i=1}^2\Phi_{2,1i}\Bigg(\sum_{j=1}^2\Phi_{1,ij}x_j\Bigg)
\end{equation}
where, $\Phi_{l,ij}$ is the learnable activation functions connecting $i^\mathrm{th}$ output node of the $l^\mathrm{th}$ layer with the $j^\mathrm{th}$ input node. Now, assuming the same example in the case of MLP, the KAN captures the nonlinear relation between inputs and output by learning the following activation functions, $\Phi_{1,1i}=[x, x]$, $\Phi_{1,2i}=[x, -x]$, and $\Phi_{2,1i}=[x^2, -x^2]$. Solving the network leads to the output $f(x)=(x_1+x_2)^2-(x_1-x_2)^2=4x_1x_2$, which is the exact relation between the inputs and the output. Unlike MLP, in KAN, we do not need to worry about making the network deeper or deciding on which nonlinear functions to use, as KAN is learning the correct nonlinear activations on its own. This mechanism of KAN allows for keeping the networks shallow and maintaining the interpretability of the input-output relation. 

\subsection{MultKAN: Multiplicative KAN}
\begin{figure}
\centering
  \includegraphics[width=1\linewidth]{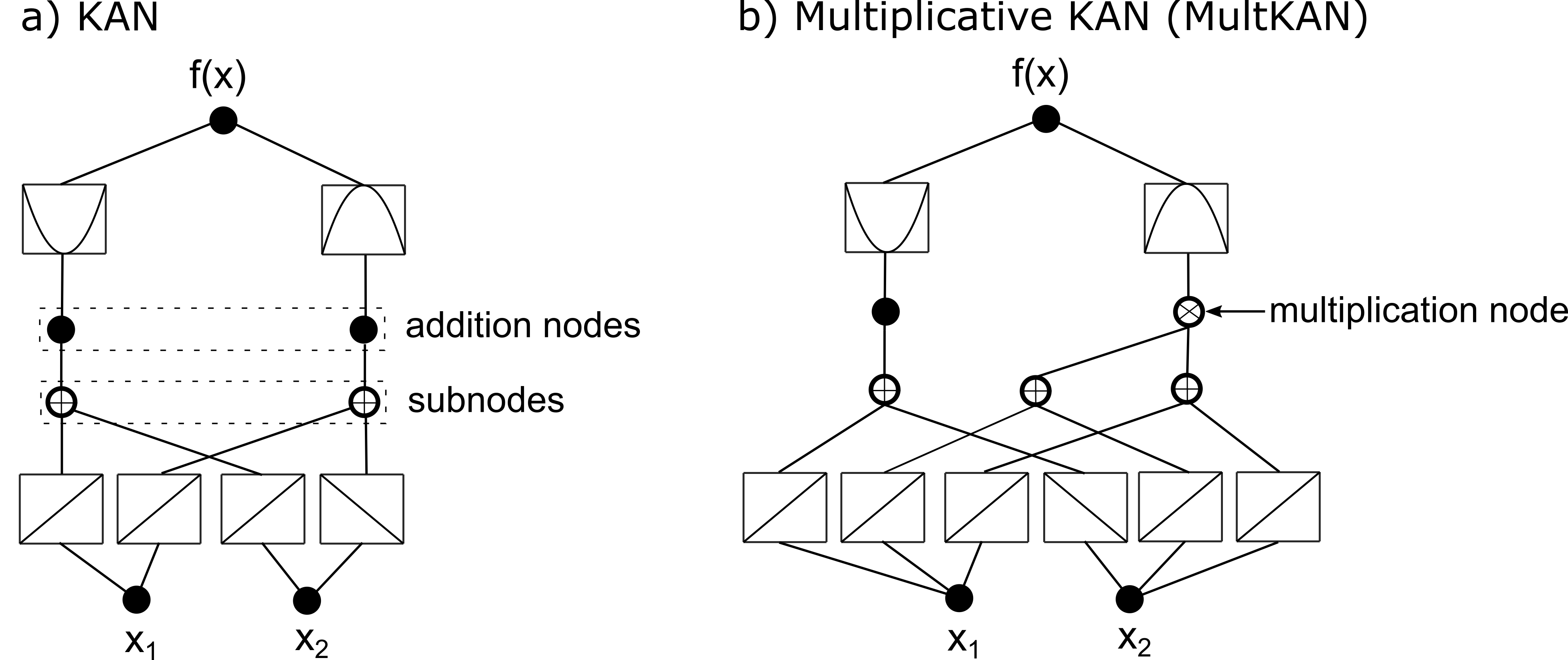}
\caption{Comparison between the KAN and MultKAN architecture}
\label{fig:KAN_MultKAN}
\end{figure}

\begin{figure}
\centering
  \includegraphics[width=0.7\linewidth]{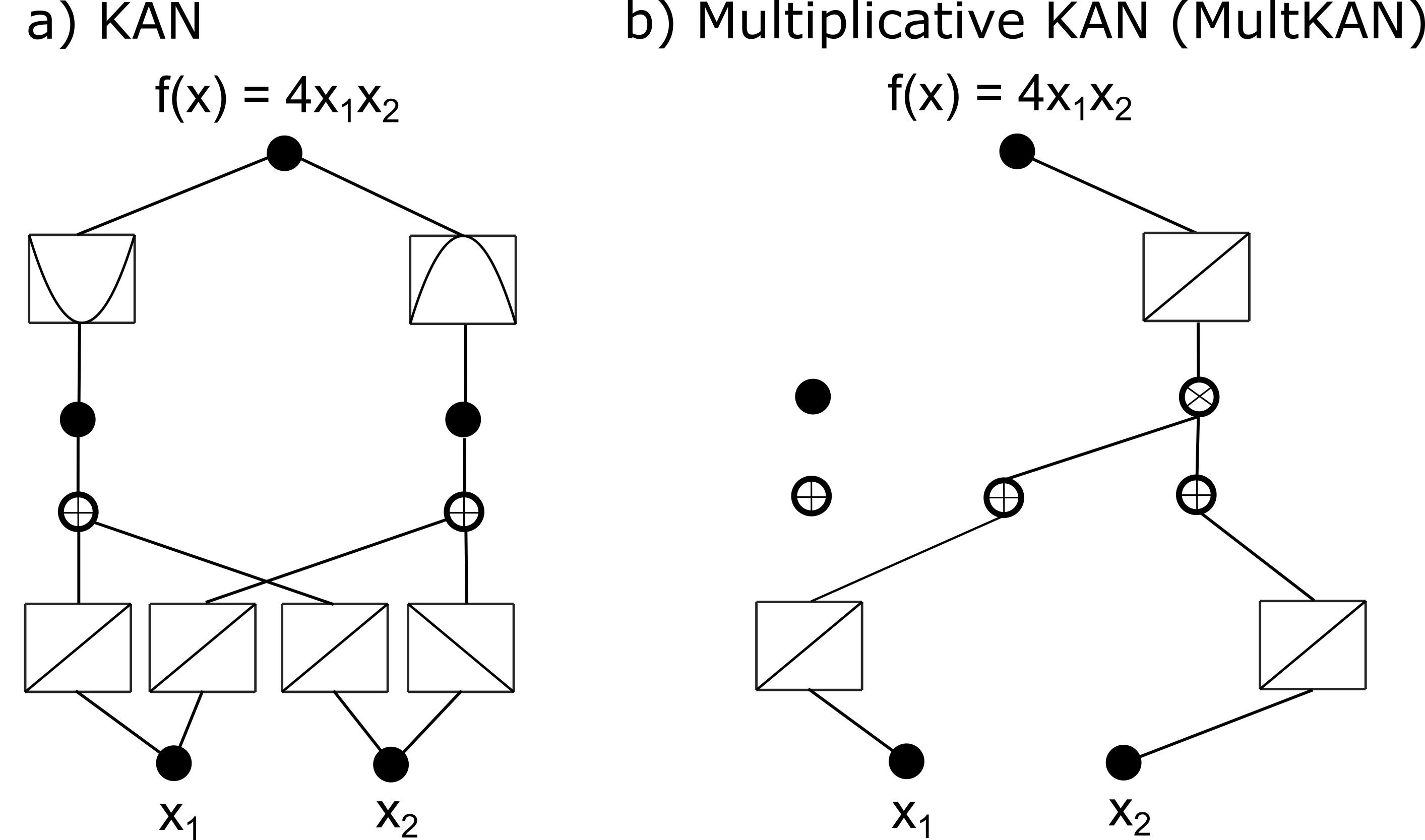}
\caption{Comparison between the interpretation of a nonlinear function from KAN and MultKAN.}
\label{fig:KAN_MultKAN_ID}
\end{figure}

Multiplicative KAN \cite{liu2024kan2} was introduced to add multiplicative operations to KAN explicitly. MultKAN is based on KAN layers with the option to allow the multiplication of output nodes of the KAN layer. The comparison between KAN and MultKAN is shown in Figure \ref{fig:KAN_MultKAN}. For distinction, the output nodes of the KAN layer are named subnodes. In KAN, the input nodes of a layer are the same as the subnodes of the previous layer. In MultKAN, some input nodes (multiplicative nodes) of a layer can be the multiplication of k-subnodes of the previous layer, while other nodes (addition nodes) are a copy of the subnodes from the previous layer. Figure \ref{fig:KAN_MultKAN} shows a MultKAN with (2,[1,1],1) nodes, where node 1 of the second layer is an addition node that is the same as the subnode 1 of the first layer, and node 2 of the second layer is a multiplicative node formed by multiplication of subnodes 2 and 3 of the first layer. The only difference between KAN and MultKAN is how the output nodes (subnodes) of a KAN layer are transformed as the input nodes of the next KAN layer.\\
This addition of multiplication operation in MultKAN offers significant flexibility in capturing the nonlinear input-output relationship, and, more importantly, it tremendously aids in maintaining the interpretability of the network. Going back to the same example used for KAN, the relationship between inputs ($x_1$, $x_2$) and the output ($4x_1x_2$) required learning six activation functions followed by the simplification of the symbolic expression. For MultKAN (see Figure \ref{fig:KAN_MultKAN_ID}), the exact relationship can be inferred by learning only three activation functions using one multiplication node without the need to simplify the symbolic expression. The output of the first layer is $x_1x_2$ based on learnable linear activation on the edges, which is further passed through linear activation in the next layer to obtain the final output. This mechanism of MultKAN aids the interpretation of the network when multiple inputs are involved, and the output contains various nonlinear terms. It is not hard to see the application of MultKAN in scientific discovery. The flexibility offered by MultKAN can be used to infer the structure of the underlying ODE/PDE and the type of nonlinearity.

\subsection{Discovery of ODE/PDE using MultKAN}
\begin{figure}[H]
\centering
  \includegraphics[width=1.0\linewidth]{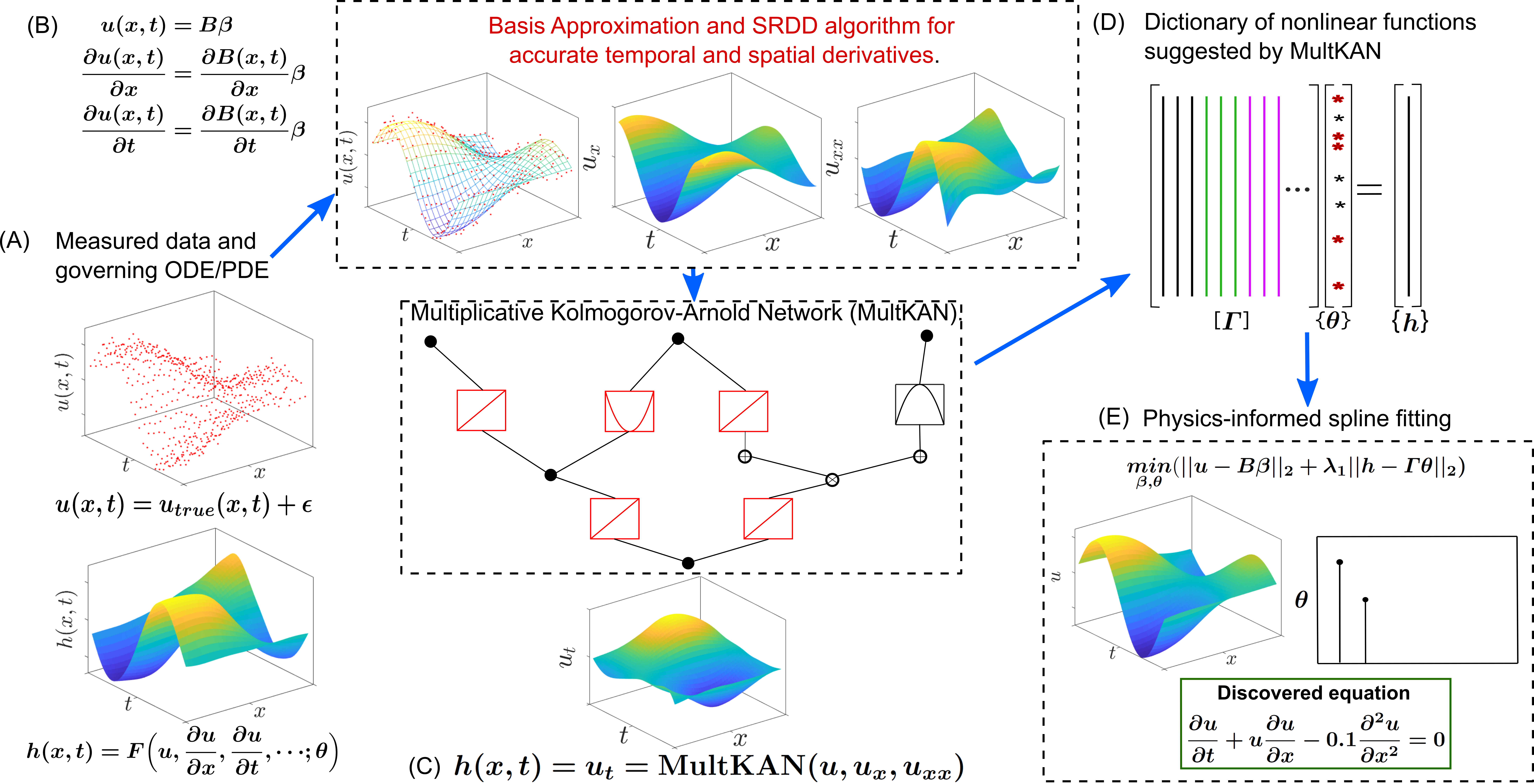}
\caption{The proposed KAN-PISF equation discovery framework: a) The measured discrete noisy data, $u(x,t)$, and the underlying unknown governing equation, $h(x,t)$, for a two-dimensional dynamic system, b) The approximation of the measured data with a linear combination of B-spline basis functions and the derivatives obtained by analytical differentiation followed by SRDD denoising, c) MulKAN architecture to obtain the approximate structure of the underlying governing equation, $h(x,t)$, d) A dictionary of functions, $\Gamma$, is created based on the functions and equation structure suggested by KAN, whose unknown coefficients, $\theta$, are found by solving the linear regression problem to approximate $h(x,t)$, and e) PISF algorithm to sequentially eliminate the excess functions from the library to converge to the true equation.}
\label{fig:idea_graphic}
\end{figure}

A differential equation can be represented in a general form as follows:
\begin{equation}
    \frac{\partial^n {u}(x,t)}{\partial t^n} = {h}(u(x,t))
\end{equation}
where, $h()$ is the nonlinear function of the $u$ and its differentials $u_t, u_x, \dots$, and so on. This study aims to discover the function $h()$ from the measured data with the help of MultKAN to understand the structure and the types of nonlinearity present in the equation. As the identification process will be developed sequentially, SRDD and PISF will be added around MultKAN to form an equation discovery algorithm that is robust to noise and applicable to various systems. The complete framework is illustrated in Figure \ref{fig:idea_graphic}. The development of the proposed algorithm is shown in the next section using an example of a forced duffing oscillator system.

\subsection{Forced duffing oscillator (noise-free)}
\begin{figure}[h]
\centering
\begin{subfigure}{.45\textwidth}
  \centering
  \includegraphics[width=0.9\linewidth]{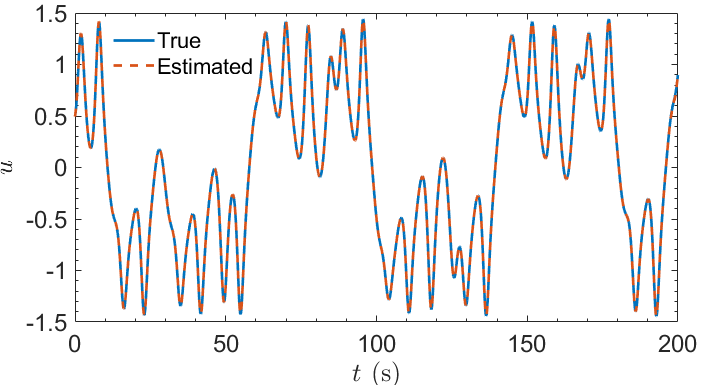}
  \caption{Displacment}
\end{subfigure}%
\begin{subfigure}{.45\textwidth}
  \centering
  \includegraphics[width=0.9\linewidth]{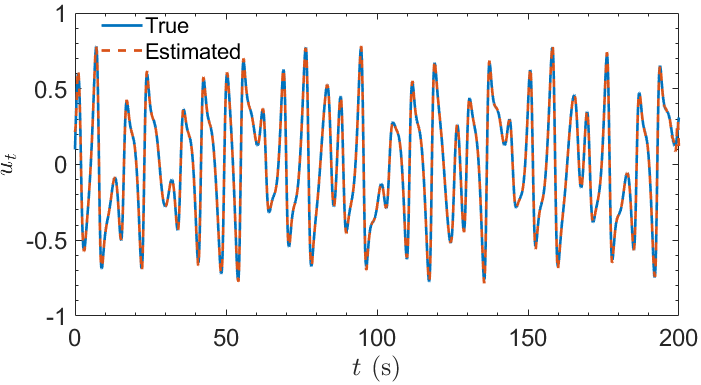}
  \caption{Velocity}
\end{subfigure}
\begin{subfigure}{.45\textwidth}
  \centering
  \includegraphics[width=0.9\linewidth]{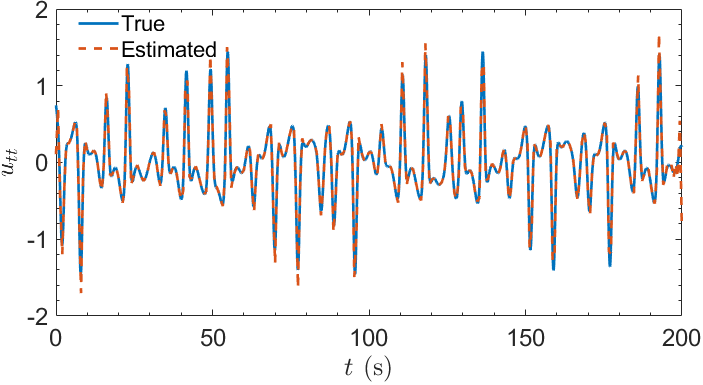}
  \caption{Acceleration}
\end{subfigure}
\caption{Comparing true and estimated displacement, velocity, and acceleration signals from KAN using automatic differentiation for noise-free data}
\label{fig:duffing_states}
\end{figure}

\begin{figure}[h]
\centering
  \includegraphics[width=1.0\linewidth]{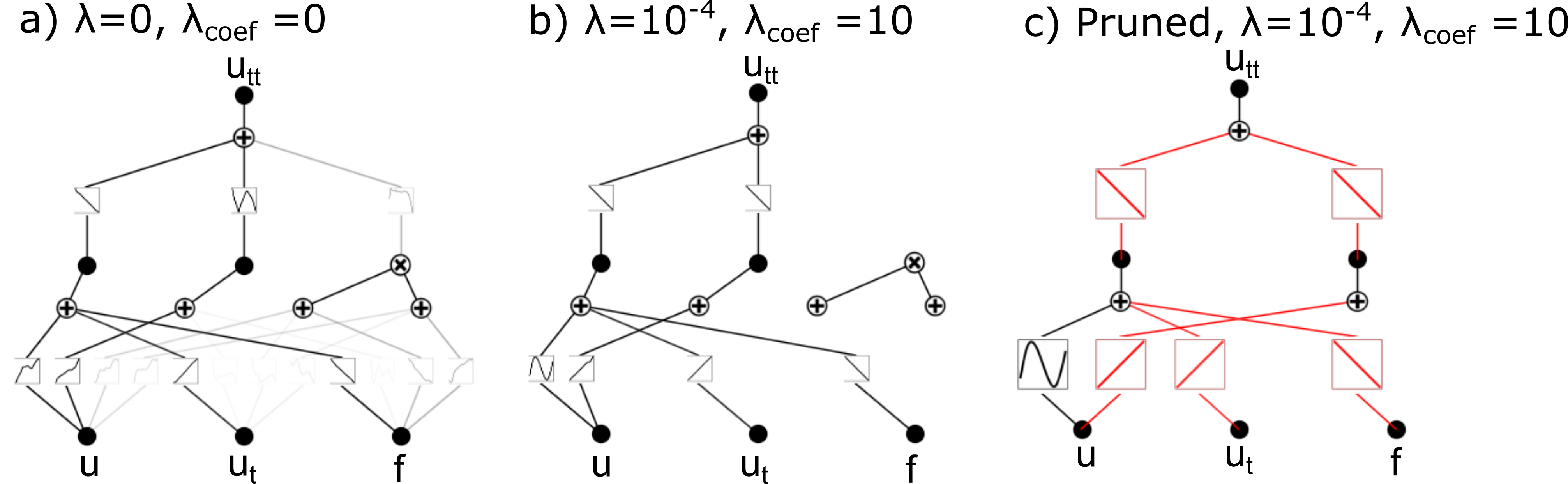}
\caption{Steps involved to converge the equation structure for the forced Duffing oscillator.}
\label{fig:duffing_0noise}
\end{figure}

\begin{figure}[h]
\centering
  \includegraphics[width=1.0\linewidth]{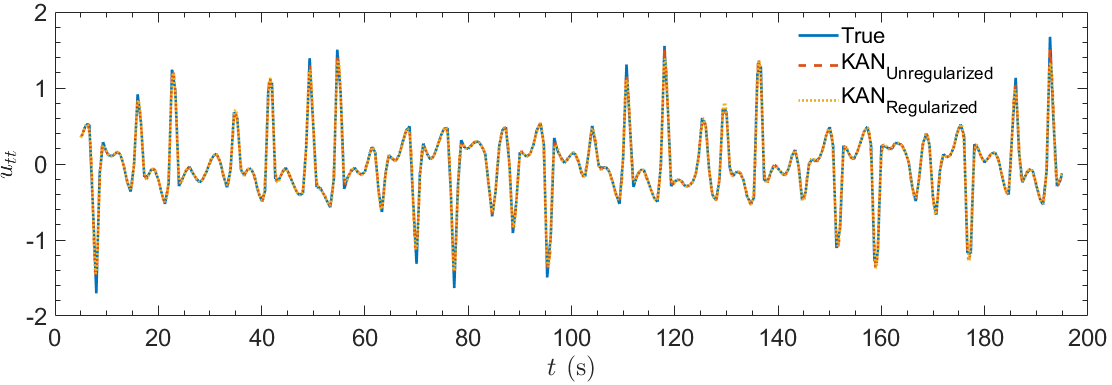}
\caption{Comparison between the output estimated by KAN with and without regularization.}
\label{fig:duffing_LHS_compare}
\end{figure}

The governing equation of the forced duffing oscillator is given by the following equation:
\begin{equation}
    u_{tt}+\theta_1u_t+\theta_2u+\theta_3u^3=\gamma cos(\omega t)
\end{equation}
where, $\theta = [\theta_1,\theta_2,\theta_3,\gamma]$ are the parameters of the equation set to [0.5,-1,1,0.42], and $\omega=1$ for the example case. The numerical solution to the equation with initial conditions $u(0)=0.5$ and $u_t(0)=0.1$ is obtained for 200 seconds at a time interval of 0.05 seconds. The measured data is assumed to be the displacement response of the oscillator. To discover the ODE, it is essential to estimate the derivatives of the measured data because, in practical conditions, all system states cannot be measured. To do this, a KAN is trained on the measured displacement signal, which takes the time signal as the input and outputs the displacement signal. Using automatic differentiation, the trained KAN is then used to generate the velocity and acceleration signals. The structure of the KAN used for this purpose is (1,1) nodes. It is a one-layer KAN with time as input and displacement as output. Figure \ref{fig:duffing_states} compares the true and estimated displacement, velocity, and acceleration signals from KAN using automatic differentiation. The estimated states are accurate and almost overlap with the true values. Automatic differentiation can produce high-quality derivatives in this case. Therefore, it can be further utilized for equation discovery.\\
A second KAN is trained to capture the nonlinear relationship between the states of the structure, as indicated in equation 4. The left-hand-side (LHS) in equation 3 is set as $u_{tt}$, and the KAN is expected to find the right-hand-side (RHS) based on equation 4. The structure of MultKAN used in this case is (3,[2,1],1) nodes. The three input nodes of the first layer correspond to $u$, $u_t$, and $f=cos(\omega t)$, the next layer has two addition nodes and one multiplication node, and the final node is the output node which is equal to $u_{tt}$. Based on equation 4, adding the multiplication node in the mid-layer is unnecessary. However, that is the whole point of equation discovery using KAN, where it should find the structure of the ODE and avoid using the multiplication node. Figure \ref{fig:duffing_0noise}a shows the equation's structure identified by KAN for the example case. It seems that KAN estimates an equation that involves numerous terms, including cross-multiplications of functions of $u$ and $f$ that do not exist in equation 4. The reason behind this is the overfitting of the data and redundancy in the functions estimating the output. Also, a few functions in the graph in Figure \ref{fig:duffing_0noise}a do not look smooth, again indicating the attempt of KAN to overfit the data to minimize the loss. The overfitting of data can be avoided using regularization that penalizes using the L1 norm of the activation functions and entropy regularization of all KAN layers (please refer to Section 2.5.2 in \cite{liu2024kan} for details) as shown:
\begin{equation}
    \Phi_{l}^{loss} = \sum_{i=1}^{n_{in}}\sum_{j=1}^{n_{out}}|\Phi_{l,ij}|_1
\end{equation}
\begin{equation}
    S^{loss}_l = -\sum_{i=1}^{n_{in}}\sum_{j=1}^{n_{out}}\frac{|\Phi_{l,ij}|_1}{\Phi_{l}^{loss}}log\Bigg(\frac{|\Phi_{l,ij}|_1}{\Phi_{l}^{loss}}\Bigg)
\end{equation}
where, $n_{in}$, $n_{out}$, $\Phi_{l}^{loss}$ and $S^{loss}_l$ are the input nodes, output nodes, L1 loss of the activation functions and entropy loss of the $l^{\mathrm{th}}$ layer, respectively. An additional regularization is applied to the coefficients of the spline function that form the activation function. It encourages splines to go to zero, proving to be effective in obtaining a sparse estimation of the activation functions. The L1 loss of the spline coefficients for the $l^{\mathrm{th}}$ layer with a grid size of $n_{grid}$ for each activation function can be written as:
\begin{equation}
    \beta_l^{loss} = \sum_{k=1}^{n_{grid}}\sum_{i=1}^{n_{in}}\sum_{j=1}^{n_{out}}|c_{l,ijk}|_1
\end{equation}
The total loss function can be expressed as the following \cite{liu2024kan}:
\begin{equation}
    loss = |y-y_{pred}|_2+\sum_{l=1}^{n_L}\lambda(\Phi_{l}^{loss}+S^{loss}_l+\lambda_{coef}\beta_l^{loss})
\end{equation}
where $y$ is the measured output, $y_{pred}$ is the predicted output, and $\lambda$ and $\lambda_{coef}$ are the hyperparameters for the overall regularization and coefficient regularization, respectively.\\
Now, applying regularization with $\lambda=10^{-4}$ and $\lambda_{coef}=10$, the structure of equation 4 is accurately identified by KAN in \ref{fig:duffing_0noise}b. The multiplication node is correctly avoided, and the activations for $u$, $u_t$, and $f$ are correctly identified as linear. The comparison of the estimated output from both unregularized and regularized networks is shown in Figure \ref{fig:duffing_LHS_compare}. Although the data fitting is almost identical for both regularized and unregularized KAN, both of these indicate an entirely different equation structure. This is evidence that regularization is absolutely necessary for discovering equations using KAN. Enforcing sparsity using regularization removes the redundancy in functions capturing the same phenomenon and encourages using fewer functions to produce the output.\\
Building on the same concept, pruning of KAN is encouraged to create as small a network as possible to increase the interpretability and avoid overfitting in some cases. Pruning is the process of removal of unused nodes and edges, such as the multiplication node and edges connecting $u_t$ and $f$ to the second subnode in Figure \ref{fig:duffing_0noise}b. The pruned network is shown in \ref{fig:duffing_0noise}c, from which the following equation structure is identified:
\begin{equation}
    u_{tt} = \theta_1u_t+\theta_2u+\theta_3F(u)+\theta_4 f
\end{equation}
where $F(u)$ is the nonlinear function of $u$ that remains to be identified. In ideal conditions, if the rest of the activations are set to identity as in Figure \ref{fig:duffing_0noise}c, the true function, $u^3$, will be the best match for $F(u)$. However, for a couple of reasons, another function might best match it: (i) due to the presence of the bias term in the KAN activation functions, (ii) sharing similar characteristics with the true function might help minimize the error. Therefore, going with the best suggestion from KAN might raise some reliability concerns and mostly lead to incorrect equations, especially for noisy data, as will be shown in upcoming results. Instead of using the best-matched function, the list of functions suggested by KAN makes the basis for forming the library of functions for the next step of the equation discovery framework. This ends the role of KAN in this study's equation discovery framework. The PISF algorithm is applied following the KAN analysis to converge to the final equation and estimate the coefficients with a high accuracy. In the next section, PISF is discussed briefly.

\subsection{Physics-Informed Spline Fitting (PISF)}
The ODE/PDE in equation 3 is written in its most general form. The dictionary-based equation discovery algorithms assume the structure of the ODE/PDE to be a linear superposition of nonlinear functions of the derivatives of the system state as shown:
\begin{equation}
    \frac{\partial^n {u}(x,t)}{\partial t^n} = \Gamma(u,u_t,u_x,\dots)\theta
\end{equation}
where, $\Gamma$ is the dictionary of functions, and $\theta$ are the coefficients of the functions. For an unknown system, since the type of nonlinearity is unknown, the dictionary $\Gamma$ is formed of a huge number of functions to cover a range of nonlinearity. The size of the dictionary makes the equation discovery difficult. The KAN plays a huge role here in suggesting the types of functions to form a small-size dictionary, which increases the probability of reaching the correct ODE/PDE.\\ 
The authors have recently proposed \cite{pal2024physics} PISF algorithm that simultaneously fits the measured data and estimates the coefficients of the library to fit the governing equation. It uses deep convolutional neural networks to optimize the following loss function:
\begin{equation}
    loss_{PISF} = ||{u}-u_{pred}||_2 + \lambda_1||h-{\Gamma\theta}||_2
\end{equation}
where, $u$ is the measured data, $u_{pred}$ is the spline fitting to the measured data, $h$ is the LHS from equation 3, and $\lambda_1$ is the hyperparameter that decides the relative importance of governing equation fitting over data fitting. In the PISF algorithm, the fitting of the data is modified simultaneously with the fitting of the governing equation. It reshapes the data and the resulting derivatives to minimize the loss in equation 11. This is unlike any other dictionary-based method where the signal and its derivatives are unchanged throughout the process. This feature of PISF helps gradually converge to the true form of the signal by decreasing the contribution of the false functions in the library. Based on the $\theta$ obtained by minimizing $loss_{PISF}$, the lowest contribution is eliminated from the library. This step is performed repeatedly until a stopping criterion is reached (refer to \cite{pal2024physics} for details), and the final equation is identified with an accurate value of the coefficients. The pseudo-code of the PISF algorithm is written below:

\begin{enumerate}
    \item Minimize $loss_{PISF}$ and find the value of the coefficients, $\theta$.
    \item Remove the function with a minimum contribution.
    \item If stopping criteria are met, STOP; else, go to step 1.
\end{enumerate}
Continuing with the example case of the forced Duffing oscillator, the library selected for PISF is $\Gamma = [u, u_t, u^2, u^3, u^4, u^5]$. This is based on the functions suggested by KAN, where the top 2 suggested functions are $u^5$ and $u^3$. The PISF algorithm successfully removes wrong functions ($u^2$, $u^4$, $u^5$) from the library and converges to the correct equation. Till now, the equation discovery was performed for noise-free data. However, in practical situations, it is impossible to get such high-quality data, and it is always contaminated with noise. The described procedure is now applied to noisy data, illustrating the need for a data-denoising algorithm.  

\subsection{Forced duffing oscillator (noisy data)}
\begin{figure}[H]
\centering
  \includegraphics[width=1.0\linewidth]{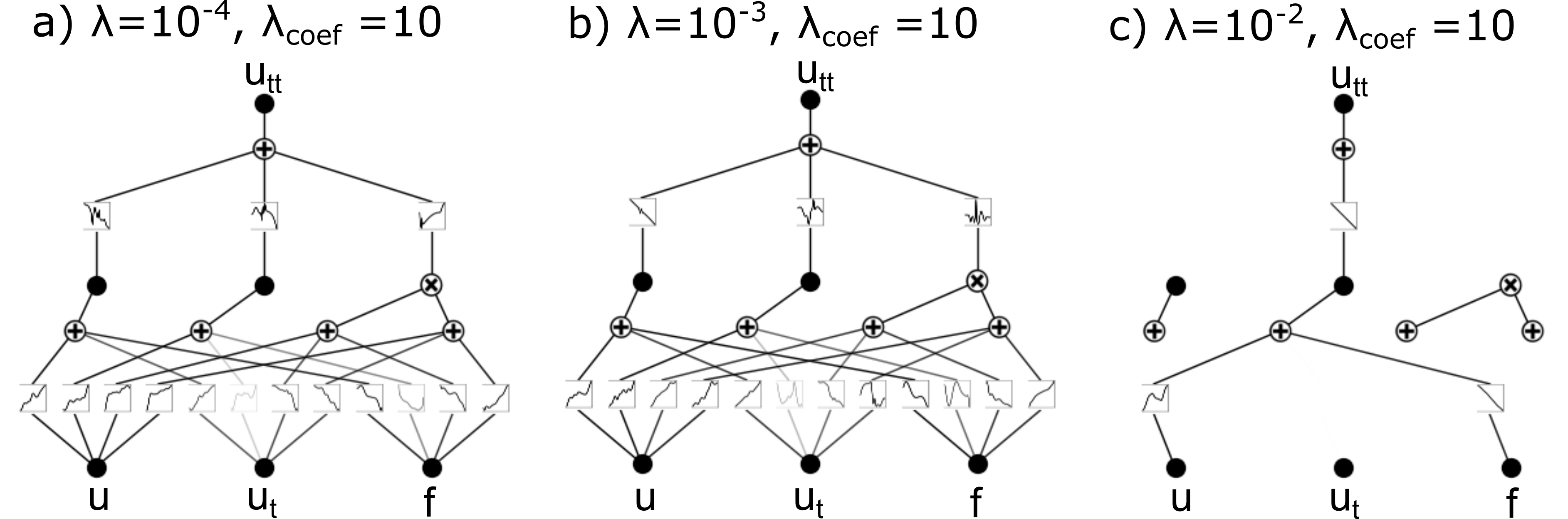}
\caption{Network structure for various levels of regularization.}
\label{fig:duffing_10noise}
\end{figure}

\begin{figure}[H]
\centering
\begin{subfigure}{.45\textwidth}
  \centering
  \includegraphics[width=0.9\linewidth]{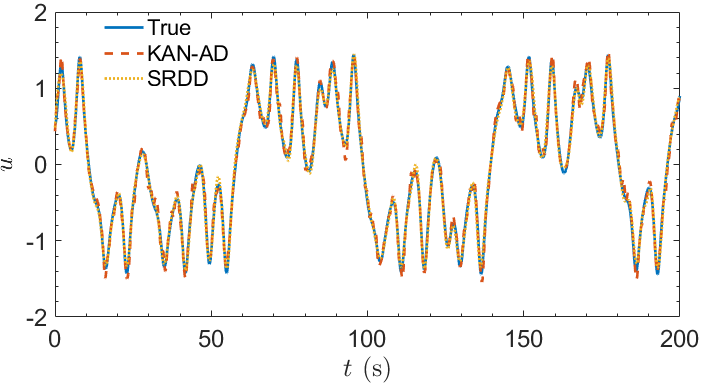}
  \caption{Displacment}
\end{subfigure}%
\begin{subfigure}{.45\textwidth}
  \centering
  \includegraphics[width=0.9\linewidth]{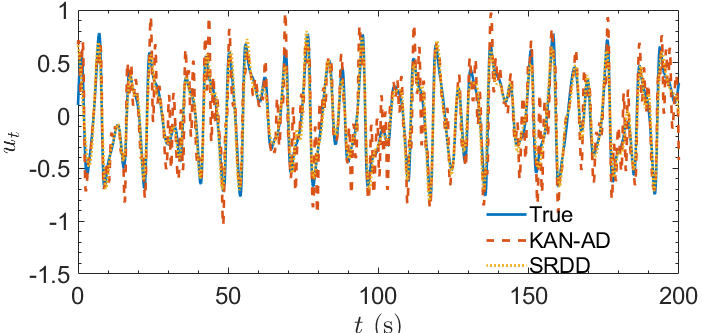}
  \caption{Velocity}
\end{subfigure}
\begin{subfigure}{.9\textwidth}
  \centering
  \includegraphics[width=0.9\linewidth]{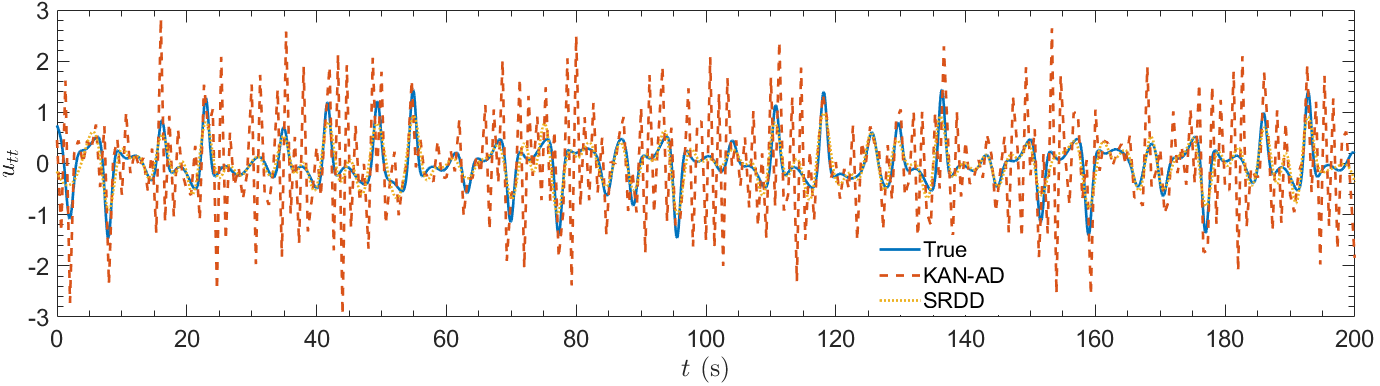}
  \caption{Acceleration}
\end{subfigure}
\caption{Comparing true and estimated displacement, velocity, and acceleration signals from KAN using automatic differentiation and SRDD for noisy data.}
\label{fig:duffing_states_10noise}
\end{figure}

The data for the forced Duffing oscillator is now contaminated with 10\% Gaussian white noise to mimic the noisy measured data. The same KAN structure and regularization values are used to identify the equation form, which was used to accomplish the equation discovery earlier successfully. Figure \ref{fig:duffing_10noise}a shows the trained KAN, which results in a dense equation structure, similar to the structure observed for the noise-free case with no regularization in Figure \ref{fig:duffing_0noise}a. Although regularization of $\lambda=10^{-4}$, $\lambda_{coef}=10$ is used in this case, the network structure is still dense, indicating overfitting. In an attempt to get a sparser estimate of the network, the regularization hyperparameter, $\lambda$ is set to $10^{-3}$, but still, the network remains dense as shown in Figure \ref{fig:duffing_10noise}b. On further increasing $\lambda$ to $10^{-2}$, the network does provide a sparse estimate. However, doing so worsens the equation fitting, i.e., the first term of equation 8 increases to compensate for the increase in $\lambda$, eventually leading to a wrong equation form as shown in Figure \ref{fig:duffing_10noise}c. To solve this problem, the source of the error must be identified that appeared after adding noise to the data.\\
The system states $u_t$ and $u_{tt}$ are derived by automatic differentiation (AD) of the output displacement from the KAN trained on the measured displacement data. Figure \ref{fig:duffing_states_10noise} shows the estimated $u_t$ and $u_{tt}$ that contain significant errors, especially for acceleration, the double derivative of displacement. When one tries to identify the equation, then erroneous velocity, erroneous displacement, and force are fitted to erroneous acceleration. Even if KAN fits the acceleration and finds a sparse solution to identify an equation, the reliability of those results is very low and likely incorrect. Therefore, data-denoising is necessary before KAN is used to identify the equation structure. The next section briefly explains the denoising algorithm.\\

\subsection{Sequentially regularized derivatives for denoising (SRDD)}
\begin{figure}[h]
\centering
  \includegraphics[width=1.0\linewidth]{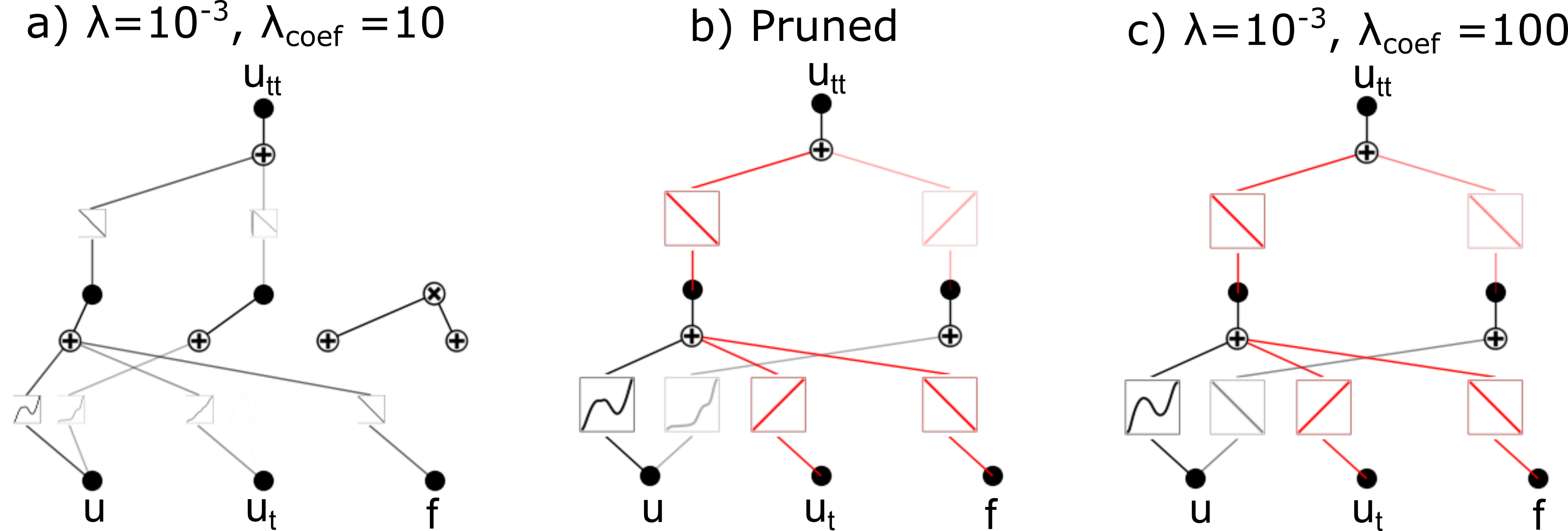}
\caption{Steps involved to converge the equation structure for the forced Duffing oscillator.}
\label{fig:duffing_SRDD_10noise}
\end{figure}

\begin{figure}[h]
\centering
  \includegraphics[width=1.0\linewidth]{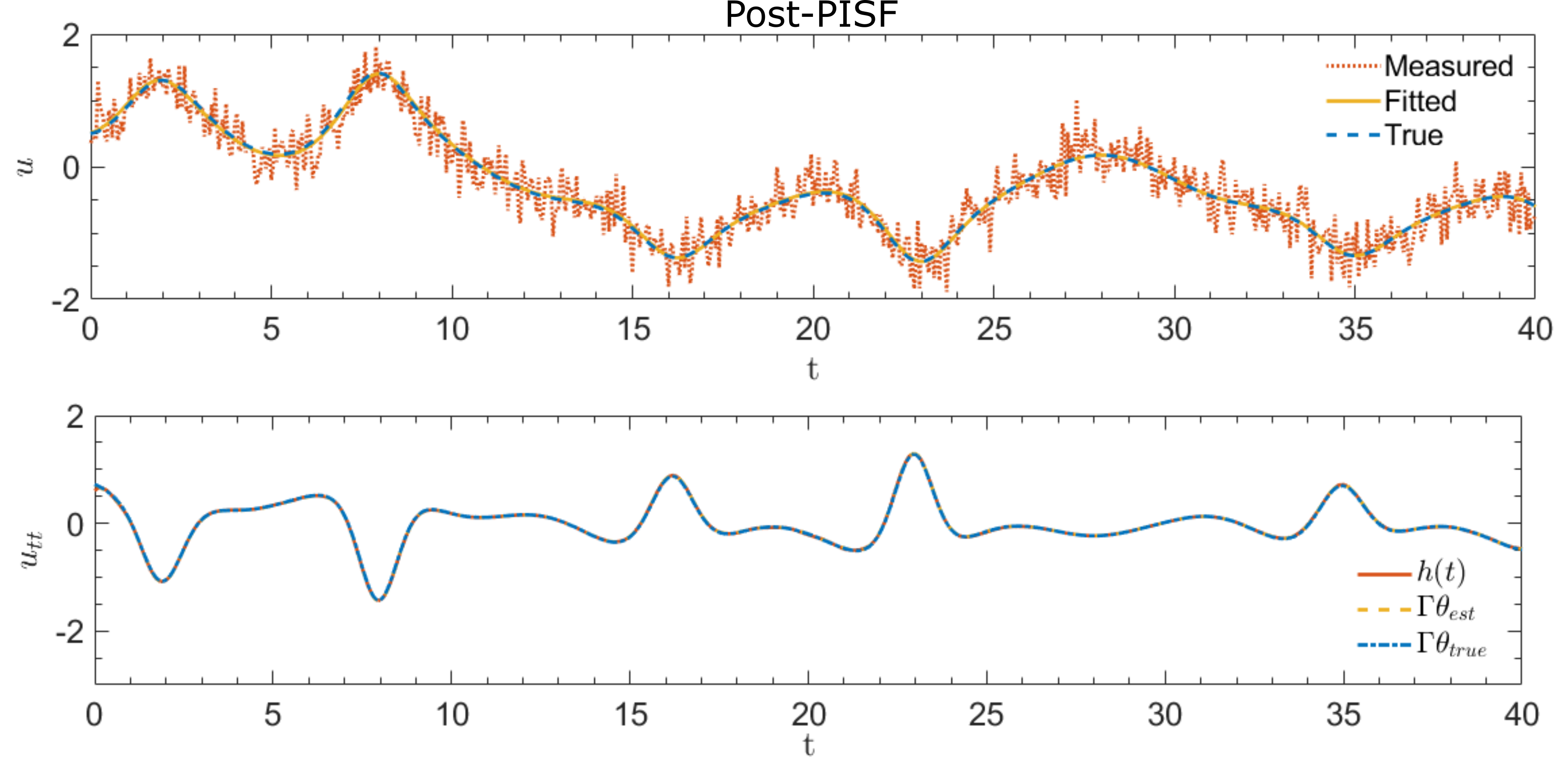}
\caption{The data and governing equation fitting post-PISF.}
\label{fig:duffing_KAN}
\end{figure}

Recently, authors proposed sequentially regularized derivatives for denoising (SRDD) algorithm \cite{pal2024physics} to obtain accurate derivatives of the measured noisy data. The algorithm is based on fitting splines to the measured data identical to 1-layer KAN with no bias in the activation function. The derivatives of the fitted data are then penalized sequentially based on the L1 norm to smooth them (please refer to \cite{pal2024physics} for details). Magnification of noise occurs when the derivatives of noisy data are numerically calculated. These errors accumulate as the order of the derivatives increases. Using splines to fit the data reduces this magnification as the derivatives can be calculated using analytical expressions of splines instead of numerical differentiation. However, if the noise is more or the splines are dense, then the magnification of the noise can still be significant for the purpose of equation discovery. The noise in the derivatives mostly occurs as unnecessary spikes that do not contribute much to the overall information in the signals. Therefore, the SRDD algorithm applies the L1 norm on the derivatives of the fitted signal sequentially, starting from a lower order derivative (say, $i_{min}$) up to a higher order derivative (say, $i_{max}$). The value of $i_{min}$ is based on the amount of smoothing needed and $i_{max}$ is based on the highest order of derivative needed (please refer to \cite{pal2024physics} for details). The following loss function is minimized during the SRDD algorithm:
\begin{equation}
    loss_{SRDD} = ||{u}-u_{pred}||_2+\lambda_2\Bigg|\frac{\partial^i {u}(t)}{\partial t^i}\Bigg|
\end{equation}
where $\lambda_2$ is the hyperparameter controlling the significance of smoothing compared to the signal fitting and $i$ is the order of the derivative that is regularized. The pseudo-code of the SRDD algorithm is given below:
\begin{enumerate}
    \item Fit the splines to the measured data.
    \item For $i$ = $i_{min}$ to $i_{max}$; minimize $loss_{SRDD}$
\end{enumerate}
The SRDD is applied to the noisy data containing 10\% noise and the derivatives obtained from the algorithm are shown in Figure \ref{fig:duffing_states_10noise}. Compared to the AD of KAN to obtain the derivatives, SRDD is doing significantly better in reducing the noise and estimating high-quality derivatives that match well with the true values. The noisy peaks in the velocity and acceleration data are completely removed because of applying regularization to them. The quality of the derivatives is now fit enough to initiate the equation discovery process.\\
KAN with the same structure as before is used to identify the equation form as shown in Figure \ref{fig:duffing_SRDD_10noise}. Using the regularization hyperparameters values $\lambda=10^{-3}$ and $\lambda_{coef}=10$, KAN correctly identifies the equation structure. It avoids the multiplication node and identifies the linear activations for $u_t$, $f$, and for the second layer, see Figure \ref{fig:duffing_SRDD_10noise}a. The next step is to prune the network to make it as small as possible, as shown in Figure \ref{fig:duffing_SRDD_10noise}b. At this stage, the linear activations are fixed and no longer learnable (marked as red). For the remaining two activation functions (marked in black), KAN is retrained to learn about them better. As shown in Figure \ref{fig:duffing_SRDD_10noise}c, one of the functions correctly converges to the linear activation, and the remaining one is the unknown nonlinear function of $u$. Similar to section 2.5, the library chosen for PISF based on KAN's suggested functions is $\Gamma = [u, u_t, u^2, u^3, u^4, u^5]$. The fitting of the measured data and governing equation post-PISF is shown in Figure \ref{fig:duffing_KAN}, and the identified equation is the following:
\begin{equation}
    u_{tt}=-0.49u_t+0.99u-1.00u^3+0.41cos(\omega t)
\end{equation}

Now the proposed framework is tested using the Van der pol oscillator and Burger's equation in the following sections.

\section{NUMERICAL RESULTS}

\subsection{Van der pol oscillator}
\begin{figure}[H]
\centering
  \includegraphics[width=1.0\linewidth]{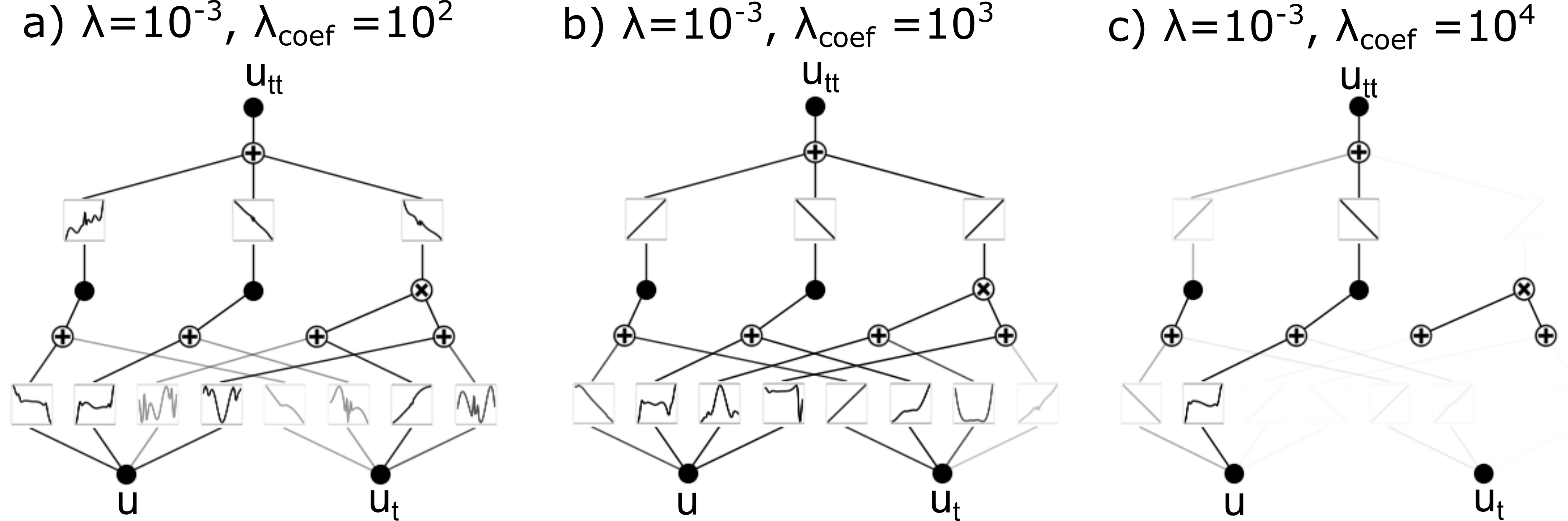}
\caption{The network structure for various values of regularization.}
\label{fig:vanderpol_10noise}
\end{figure}

\begin{figure}[H]
\centering
\begin{subfigure}{.45\textwidth}
  \centering
  \includegraphics[width=0.9\linewidth]{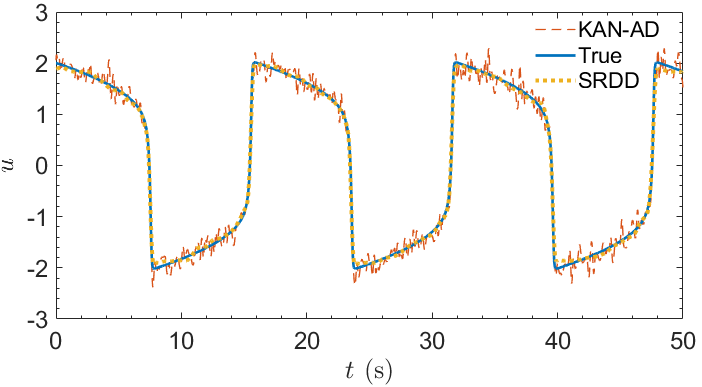}
  \caption{Displacment}
\end{subfigure}%
\begin{subfigure}{.45\textwidth}
  \centering
  \includegraphics[width=0.9\linewidth]{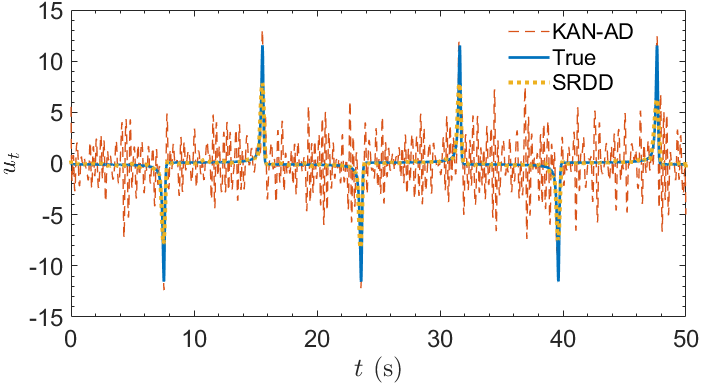}
  \caption{Velocity}
\end{subfigure}
\begin{subfigure}{.9\textwidth}
  \centering
  \includegraphics[width=0.9\linewidth]{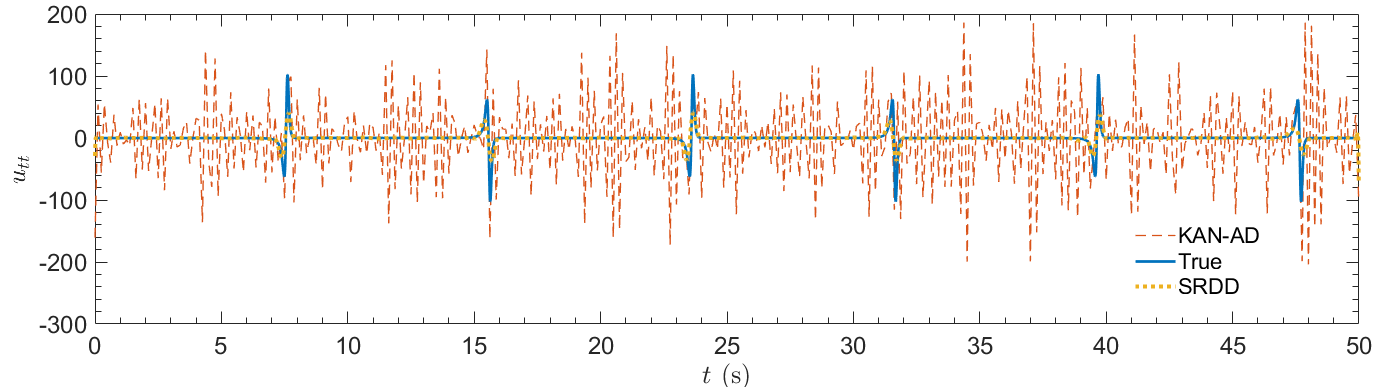}
  \caption{Acceleration}
\end{subfigure}
\caption{Comparing true and estimated displacement, velocity, and acceleration signals from KAN using automatic differentiation for noisy data}
\label{fig:vanderpol_states_10noise}
\end{figure}

The governing equation of the Van der Pol oscillator is the following:
\begin{equation}
    u_{tt}+\theta_1u_t+\theta_2u^2u_t+\theta_3u=0
\end{equation}
where, $\theta = [\theta_1, \theta_2, \theta_3]$ are the parameters corresponding to linear damping, nonlinear damping, and stiffness, respectively. This equation for parameter choice of $\theta = [-8, 8, 1]$ is a stiff-differential equation, and it poses significant challenges for equation discovery in the presence of noise. The ODE was solved for displacement response, $u$, for 50 seconds at a sampling time of 0.01 seconds. To mimic the experimental conditions, 10\% Gaussian noise is added to the displacement data. For the sake of illustration, equation discovery is first attempted using the velocity and acceleration signal obtained by AD. For this, KAN with a structure (1,1) nodes is trained on the displacement data using time as the input, and AD of displacement from KAN is used to obtain the velocity and acceleration data. Another KAN with a structure (2,[2,1],1) with displacement and velocity as inputs and acceleration as the output is used for equation discovery. Figure \ref{fig:vanderpol_10noise} shows the network structure for various regularization values. For relatively lower regularization values, the KAN structure remains dense, and the activation function shapes are also jagged and difficult to interpret. For higher regularization value, the network structure is sparse but definitely incorrect. The reason for such behavior is again related to the velocity and acceleration signals obtained from AD of the displacement from KAN. The fitted displacement and calculated velocity and acceleration signal are compared in Figure \ref{fig:vanderpol_states_10noise}. The velocity and acceleration signals obtained from AD show significant noise magnification from the noisy displacement fitting in Figure \ref{fig:vanderpol_states_10noise}a. Rapid changes in the system state that occur periodically should be captured in the estimated signals to learn the system's behavior and derive the correct equation of motion. These rapids are clearly seen in the displacement and fairly visible in the estimated velocity. However, the noise in the estimated acceleration is the same or higher magnitude than the acceleration near the rapids. It is unlikely that the KAN network will converge to the correct form of the equation, given the quality of signals produced. This is why regularization does not help converge to the correct sparse form of the network in Figure \ref{fig:vanderpol_10noise}. This example again illustrates that relying on AD to obtain the velocity and acceleration response is unreliable and low-quality for equation discovery. 

\begin{figure}[H]
\centering
  \includegraphics[width=1.0\linewidth]{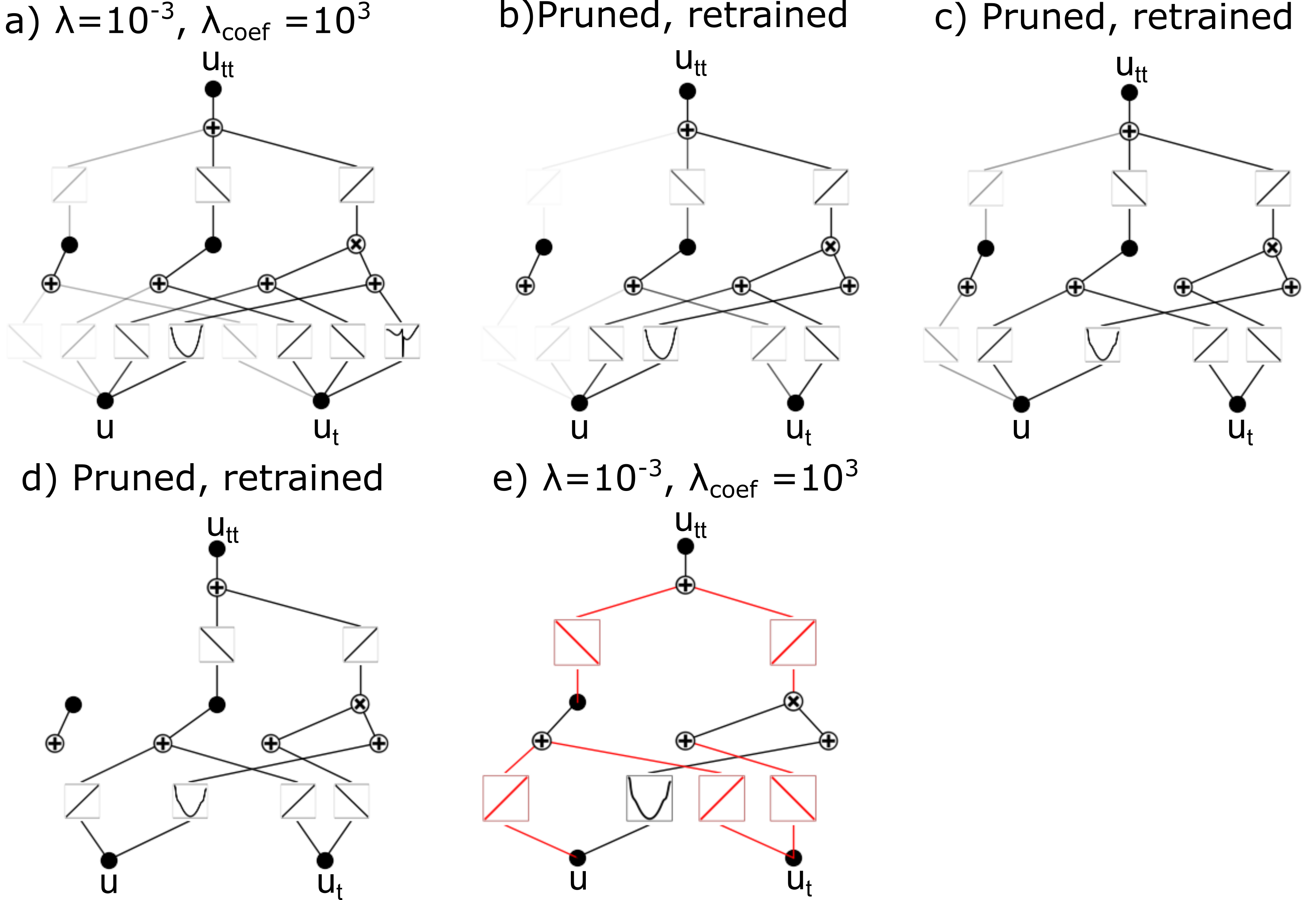}
\caption{Steps involved to converge to the equation structure for the Van der Pol oscillator.}
\label{fig:vanderpol_SRDD}
\end{figure}

\begin{figure}[H]
\centering
  \includegraphics[width=1.0\linewidth]{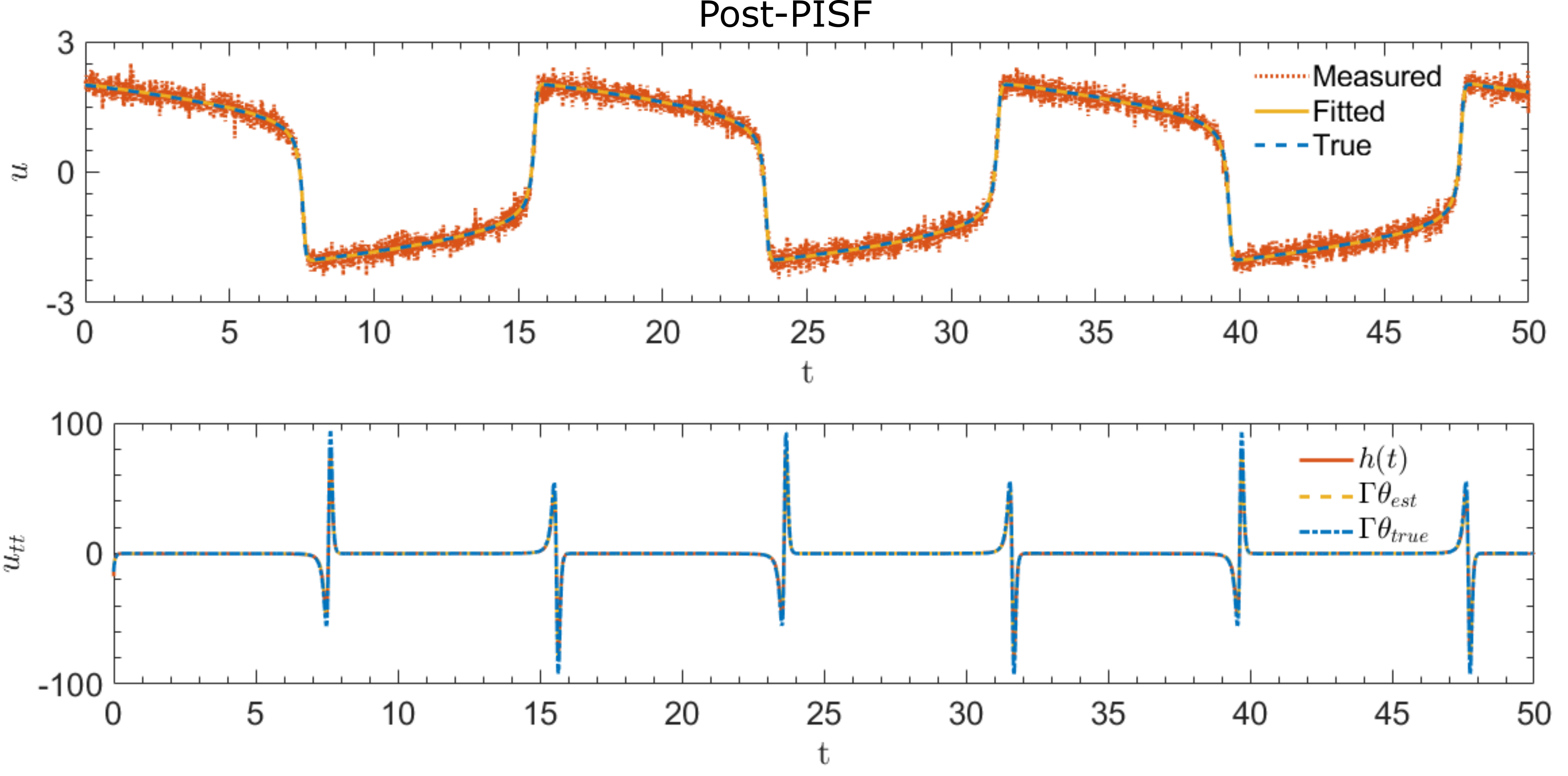}
\caption{The data and governing equation fitting post-PISF for Van der Pol oscillator.}
\label{fig:vanderpol_KAN}
\end{figure}

The denoising of the signals is conducted using the SRDD algorithm, producing high-quality displacement, velocity, and acceleration, as shown in Figure \ref{fig:vanderpol_states_10noise}. The unnecessary noisy peaks seen between the rapids in the velocity and acceleration signals obtained from AD are absent in the case of SRDD algorithm. The velocity and acceleration do not show any false features during the slow-variation region of the signal but the high velocity and acceleration changes are captured accurately at the rapids. The signals obtained from the SRDD algorithm are used for equation discovery using KAN.\\
Figure \ref{fig:vanderpol_SRDD} shows the steps to reach the correct form of the equation. In the first run, regularization values are set to $\lambda=10^{-3}$ and $\lambda_{coef}=10^3$. The obtained network is dense but unlike Figure \ref{fig:vanderpol_10noise}, there seems to be redundancy that can be easily inferred. For instance, the activations for $u$ and $u_t$ connecting to the first two subnodes are linear for both, and the further activations in the next layer are also linear. Implying that one of the addition nodes is redundant. Also, the activation between $u_t$ and the fourth subnode is strange, mostly seen when some overfitting occurs. The network is pruned in steps, starting with the removal of the edge connecting $u_t$ with the fourth subnode. The network is retrained, and Figure \ref{fig:vanderpol_SRDD}b shows the obtained network structure. Next, pruning is carried out by removing the edge connecting $u$ with the third subnode. The logic behind this pruning is that the multiplication node is used to multiply two functions of $u$. Removing one of the functions from the multiplication node does not impact the fitting, as the removed nonlinearity can be captured with the two available addition nodes. Since both of the addition nodes are linear activations, removing one activation from the multiplication node should change one of the activations of the addition node to the desired nonlinear function. Figure \ref{fig:vanderpol_SRDD}c shows the resultant network structure, which does not show any change to the activations of the addition node. Next, pruning is carried out by removing the linear activation from $u$ as it is redundant, and the retrained network is shown in Figure \ref{fig:vanderpol_SRDD}d. Since no more pruning can be performed, the linear activations are fixed (marked as red), and the network is retrained to obtain the suggested nonlinear functions for the remaining activation. The equation structure inferred from Figure \ref{fig:vanderpol_SRDD}e has the following form:
\begin{equation}
    u_{tt}=\theta_1u_t+\theta_2F(u)u_t+\theta_3u
\end{equation}
where, $F(u)$ is a nonlinear function of $u$. The top two suggested functions by KAN are $u^4$ and $u^2$, out of which $u^2$ is the true function. Based on the suggestions, the following library, $\Gamma = [u, u_t, u^2u_t, u^3u_t, u^4u_t]$, is used for the next step of PISF. The authors would like to point out that the correct equation form is interpreted in Figure \ref{fig:vanderpol_SRDD}c itself. However, two extra steps are carried out to shrink the network to avoid any form of overfitting due to the bias terms in the activation functions so that more reliable suggestions for the nonlinear terms are obtained from KAN. The fitting of the measured data and governing equation post-PISF is shown in Figure \ref{fig:vanderpol_KAN}, and the identified equation is the following:
\begin{equation}
    u_{tt}=7.23u_t-7.40u^2u_t-0.92u
\end{equation}

\subsection{Burger's equation}
\begin{figure}[ht]
\centering
  \includegraphics[width=0.6\linewidth]{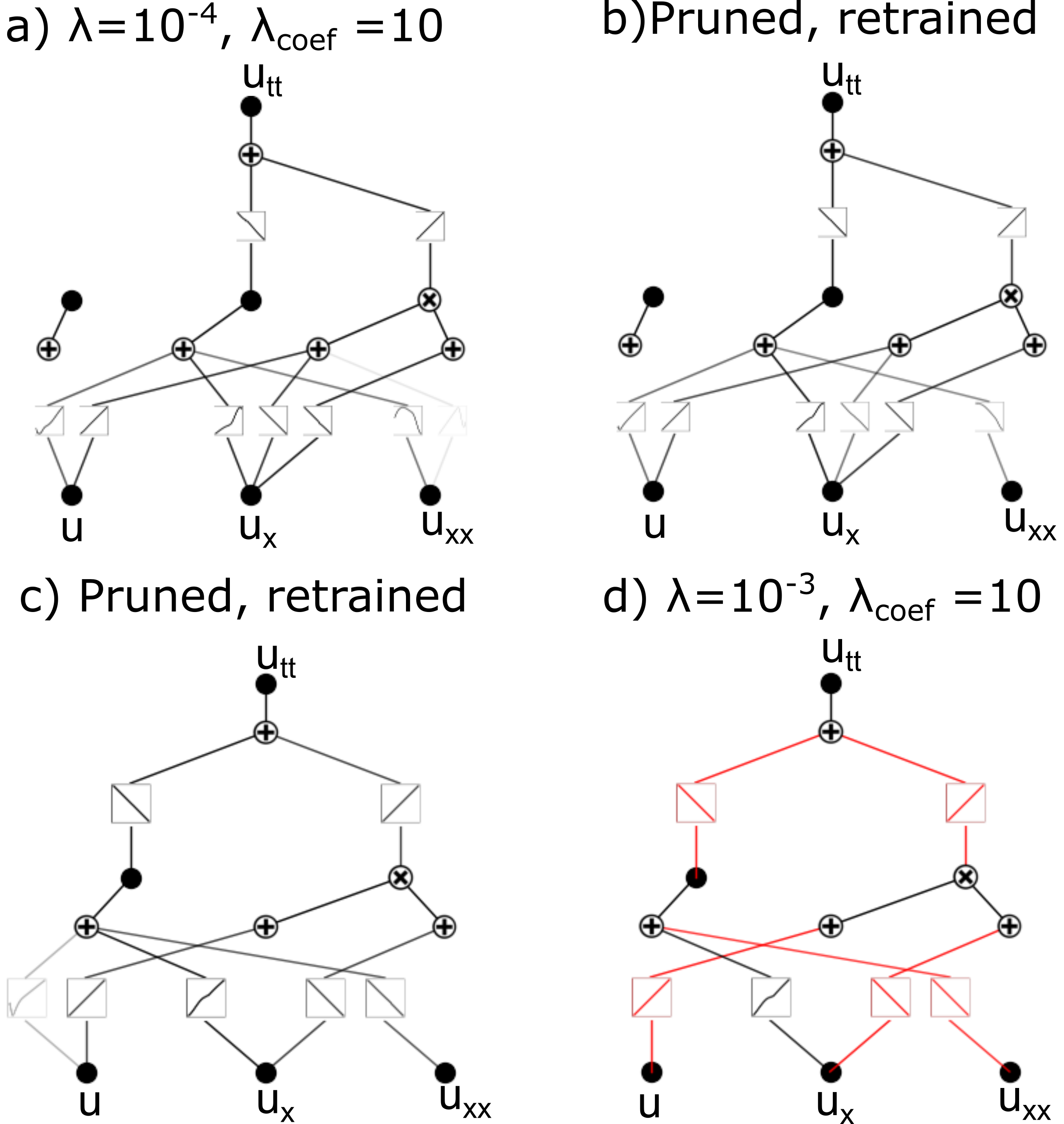}
\caption{Steps involved to converge the equation structure for the Burger's equation}
\label{fig:burgers_SRDD}
\end{figure}

\begin{figure}[ht]
\centering
  \includegraphics[width=1.0\linewidth]{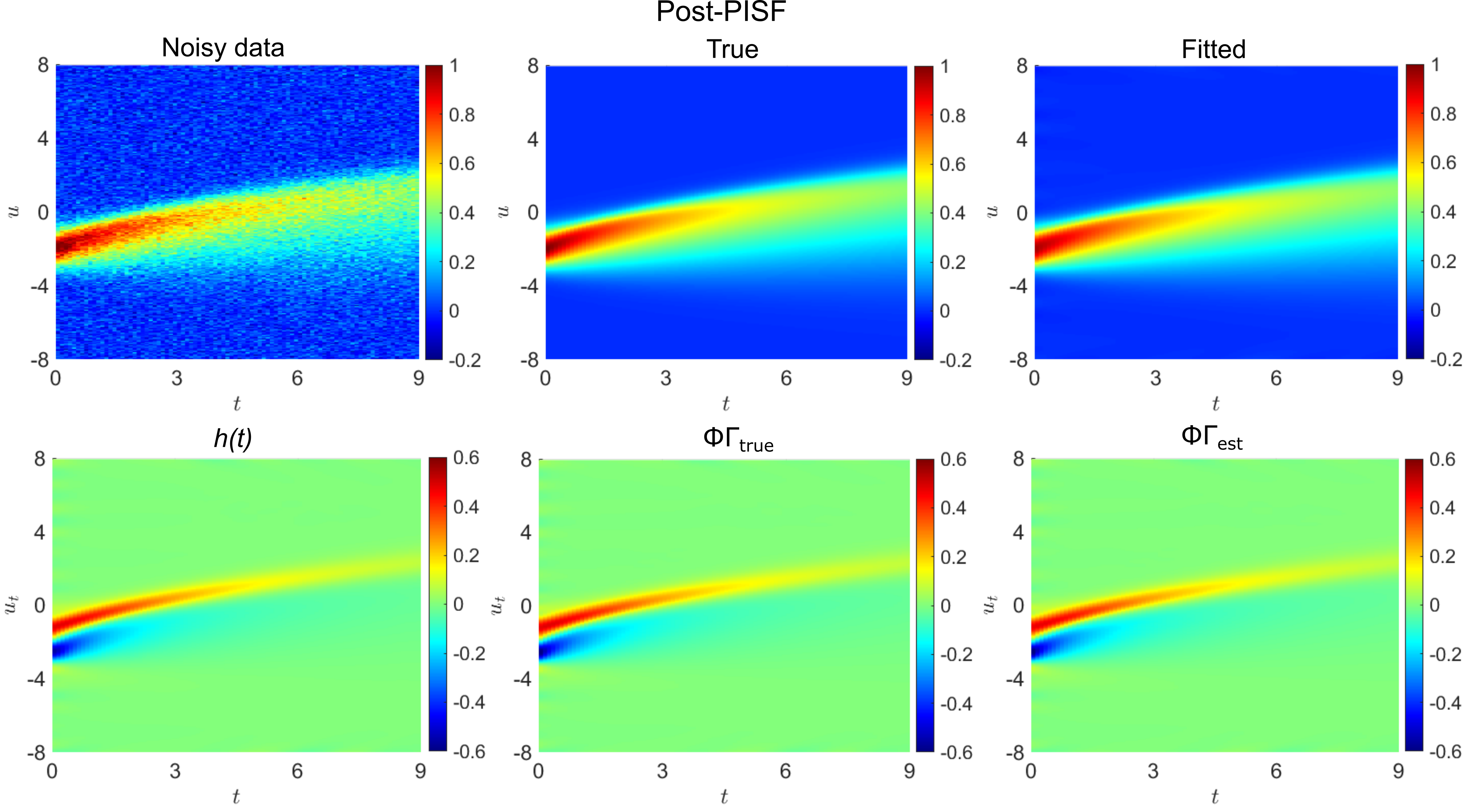}
\caption{The data and governing equation fitting post-PISF for the Burger's equation.}
\label{fig:burgers_KAN}
\end{figure}

The governing PDE of the Burger's equation has the following form:
\begin{equation}
    u_t+\theta_1uu_x+\theta_2u_{xx}=0
\end{equation}
where $\theta=[\theta_1,\theta_2]$ are the parameters of the equation related to the viscosity of the fluid. The PDE was solved for the displacement response $u \in \mathbb{R}^{256 \times 96}$ \cite{rudy2017data} for $\theta=[1,-0.1]$, to which 10\% noise is added to mimic measured data. The data is first denoised using the SRDD algorithm, and then a KAN network is trained with the structure of [3,[2,1],1] nodes. The three inputs are $u$, $u_x$, and $u_{xx}$ and the output is $u_t$. Figure \ref{fig:burgers_SRDD} shows steps to reach the equation form using KAN. Beginning training with regularization values $\lambda=10^{-4}$ and $\lambda_{coef}=10$, the network provides fairly sparse estimate. In the next step, the network is pruned by removing the edge connecting $u_{xx}$ with the third subnode and retrained to get the form in Figure \ref{fig:burgers_SRDD}b. In this step, the multiplication node shows the multiplication of activations functions of $u_x$ even when one addition node is left unused. Therefore, one linear activation going to the multiplication node is removed for $u_x$, and retraining is performed. The activation function seems redundant because the addition node is still unused and removed during pruning in Figure \ref{fig:burgers_SRDD}c. At this point, the activation functions from $u$ and $u_x$ to the first subnode show nonlinear activations. The linear activation functions are fixed (shown in red in Figure \ref{fig:burgers_SRDD}d) to yield the final network structure. After retraining, the edge connecting $u$ with the first subnode was pruned due to little contribution, while the same couldn't be done for $u_x$. The final structure of the equation obtained from KAN is the following:

\begin{equation}
    u_t=\theta_1uu_x+\theta_2u_{xx}+F(u_x)
\end{equation}
The identified equation structure has the additional term $u_x$ compared to the true form in equation 18. For the final step of PISF, the chosen library is $\Gamma = [u_x, u_{xx}, u_x^2, uu_x]$. The PISF can identify the correct equation form by eliminating $u_x$ and $u_x^2$. The data and PDE fitting by PISF are shown in Figure \ref{fig:burgers_KAN}, and the final identified equation is the following:
\begin{equation}
    u_t=-0.99uu_x+0.1u_{xx}
\end{equation}
The authors would like to mention here that the structure of Burger's equation can be found exactly in just one training if noise-free data is used. The presence of noise causes several problems and poses difficulties in finding a sparse solution. Therefore, the pruning of KAN is carried out in several steps before deriving the equation form for noisy data. In the example cases, 10\% noise is used, which is a significant amount. The amount of pruning steps would be reduced for lower noise levels, and sparse forms can be easily found.

\subsection{Bouc-Wen Hysteresis model}
\begin{figure}[H]
\centering
  \includegraphics[width=1.0\linewidth]{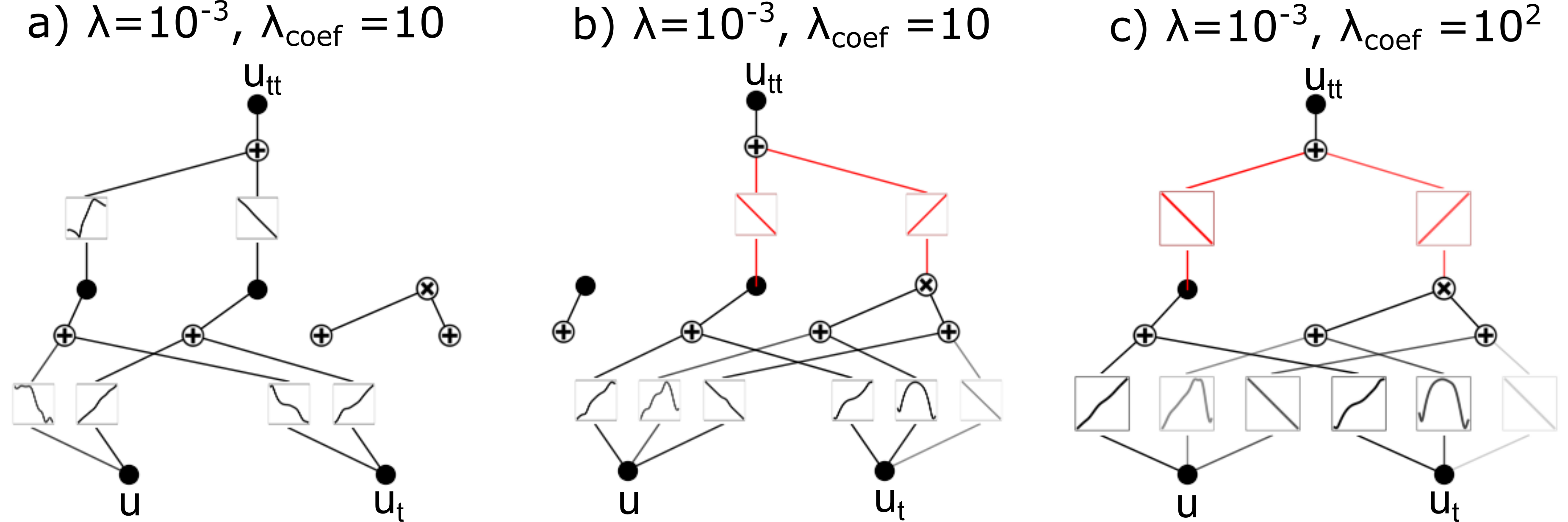}
\caption{Steps involved to converge to the equation structure for approximating the Bouc-Wen model.}
\label{fig:hysteretic_0noise}
\end{figure}

\begin{figure}[H]
\centering
  \includegraphics[width=0.8\linewidth]{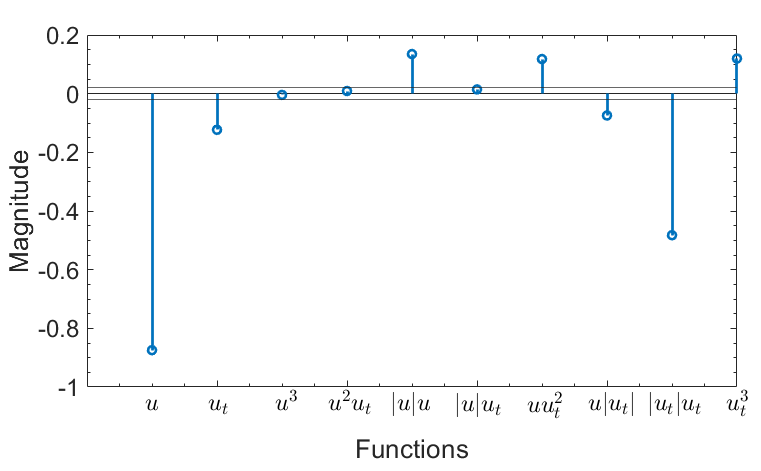}
\caption{The coefficient values for the functions in the library. The thresholding band shows the functions falling within it will be removed from the library.}
\label{fig:hysteretic_thresh}
\end{figure}

The Bouc-Wen model is used to describe the equation of motion of nonlinear hysteretic systems with a cosine forcing function as follows:
\begin{equation}
    mu_{tt}+cu_t+\alpha ku+(1-\alpha)kz=2cos(\omega t)
\end{equation}
\begin{equation}
    z_t = u_t\{A-[\beta \mathrm{sign}(zu_t)+\gamma]|z|^n\}
\end{equation}
where $m$, $c$, $k$, and $z$ are the system's mass, damping, linear stiffness, and the unobservable hysteretic displacement. $A$, $\beta$, $\gamma$, and $n$ are the Bouc-Wen parameters controlling the model's behavior. The coupled ODE is solved for $m=1$, $c=0.2$, $k=1$, $\alpha=0.2$, $A=2$, $\beta=1$, $\gamma=0.5$, $n=2$, and $\omega=1$ to obtain the displacement, velocity, and acceleration response of the system for 20 seconds long with a sampling time of 0.02 seconds. We use noise-less data for this system and will not need SRDD for denoising and PISF to filter functions because no update to signal fitting is required as the signals are accurate. The filtering of functions in the library is achieved by thresholding.\\
The KAN has the structure of (2,[2,1],1) nodes, with $u$ and $u_t$ as inputs and $u_{tt}$ as output. The variable $z$ cannot be used as it is an unobservable quantity. Therefore, we will always obtain an approximate model of the system's behavior. Figure \ref{fig:hysteretic_0noise} shows the steps to reach the form of the equation. Although a sparse form of the equation is obtained in Figure \ref{fig:hysteretic_0noise}a, the interpretation is difficult as most activation functions do not correlate well with the list of functions. This is probably the result of overfitting the data as we are trying to model a system of equations with a single ODE. To avoid this, the activations of the second layer are set to linear, and the model is retrained to obtain the network as shown in Figure \ref{fig:hysteretic_0noise}b. This time, a sparse network is achieved with interpretable functions. Further pruning and retraining are performed to get the final form and shape of the activation functions. Based on the suggested functions from KAN and network structure, the library created is $\Gamma = [u,u_t,u^3,u^2u_t,|u|u,|u|u_t,u_t^2u,|u_t|u,|u_t|u_t,u_t^3]$. As described in equation 10, the system of equations is solved to get the coefficients of the functions in the library. The low contributing nonlinear functions are thresholded (see Figure \ref{fig:hysteretic_thresh}), and the remaining functions form the identified equation of motion.  The identified approximate ODE for the nonlinear hysteric system is the following:

\begin{equation}
    u_{tt}=-0.855u-0.068u_t+0.114|u|u+0.109uu_t^2-0.056|u_t|u-0.437|u_t|u_t+0.095u_t^3
\end{equation}

\begin{figure}[H]
\centering
\begin{subfigure}{.45\textwidth}
  \centering
  \includegraphics[width=0.9\linewidth]{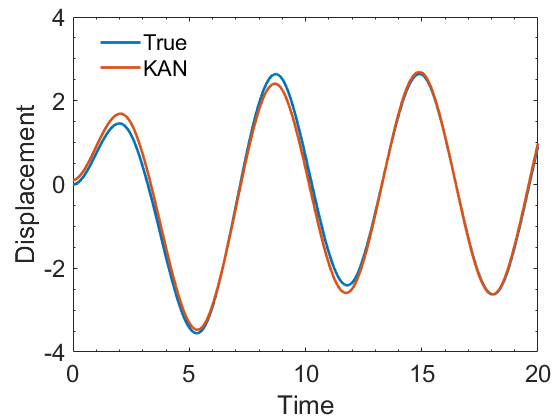}
  \caption{Displacement}
\end{subfigure}%
\begin{subfigure}{.45\textwidth}
  \centering
  \includegraphics[width=0.9\linewidth]{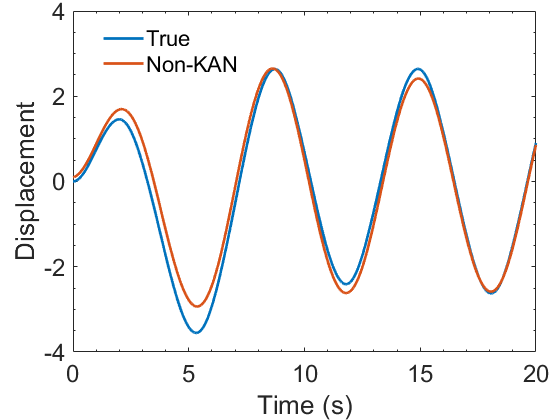}
  \caption{Displacement}
\end{subfigure}
\begin{subfigure}{.45\textwidth}
  \centering
  \includegraphics[width=0.9\linewidth]{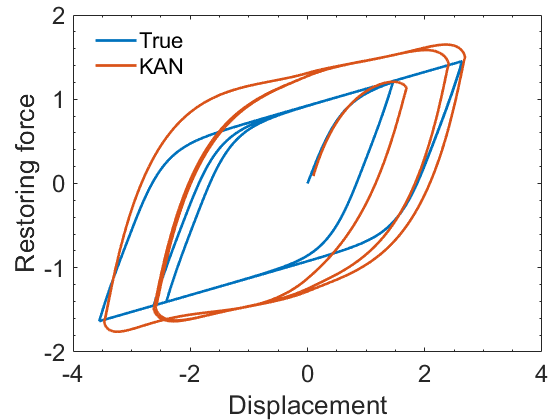}
  \caption{Hysteresis}
\end{subfigure}%
\begin{subfigure}{.45\textwidth}
  \centering
  \includegraphics[width=0.9\linewidth]{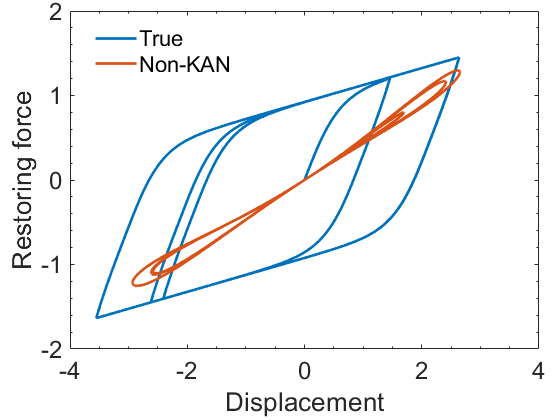}
  \caption{Hysteresis}
\end{subfigure}
\caption{Comparing displacement and restoring force (hysteresis plot) for the model identified from the library suggested by KAN (a,c) with an identified model from the library formed by common functions (b,d).}
\label{fig:hysteresis_compare}
\end{figure}

The identified equation contains modulus functions of $u$ and $u_t$, which KAN suggested. Modulus functions are an uncommon category of functions and are not used for equation discovery in studies using a dictionary of functions. The commonly used functions are the polynomials and cross-multiplication of polynomials with derivative functions. Assuming if we were identifying this system using the commonly used dictionary of functions $\Gamma = [u,u_t,u^2,u^3,u^4,u^5,uu_t,u^2u_t,u^3u_t]$, we converge to the following equation:

\begin{equation}
    u_{tt} = -0.45u-0.59u_t-0.012u^2+0.013u^2u_t
\end{equation}
To compare the two identified equations, 22 and 23, the response is simulated for the same cosine forcing function. The displacement responses and the restoring forces from both equations are shown in Figure \ref{fig:hysteresis_compare}. The simulated displacement response from both equations shows a similar quality, with KAN doing slightly better in the first half of the signal. At the same time, the shape of the restoring force from the two equations is completely different. The KAN-based equation captured the hysteresis significantly better than the non-KAN-based equation, which fails to capture the hysteresis curve and is mostly linear. This illustrates the advantage of using KAN, which can suggest relevant functions from a bigger pool of functions while maintaining a low-complexity network. If the complete pool of functions is used in the library for existing methods, then it becomes overly complex and difficult to obtain a sparse solution.\\

\section{CONCLUSIONS}
In this study, an equation discovery framework is proposed based on the Kolmogorov-Arnold network accompanied by SRDD and PISF algorithms to ensure robustness against noise and generate reliable equations. KAN performs the critical task of identifying the structure of the equation and suggesting the relevant nonlinear functions from a large pool of functions. This feature of KAN allows the scanning of a large number of functions while maintaining a low-complexity network for interpretability. In the presence of noise, calculating high-quality derivatives from the measured data is necessary to derive the governing ODE/PDE. It was seen that erroneous derivatives lead to physically meaningless signals with no hope of converging to the correct equation. The SRDD algorithm could denoise the data to obtain accurate derivatives fit for equation discovery, compared to automatic differentiation that produced erroneous signals. KAN's suggested functions and the network structure are used to create a small overcomplete library of functions. This library is used in the PISF algorithm, which filters out the excess functions based on function contributions to converge to the correct equation.\\
The framework was illustrated on the forced Duffing oscillator, Van der Pol oscillator, and Burger's equation. The SRDD algorithm generated accurate derivatives compared to automatic differentiation, especially in the Van der Pol oscillator (a stiff ODE), which experiences a huge magnification of error in the derivatives. The regularization of KAN proved to be critical to identifying the correct form of the equation structure. If insufficient regularization is used, then KAN suggests a dense network that is overfitted and difficult to interpret. Pruning the network to shrink its size is helpful in identifying the equation structure and improving the quality of functions suggested by KAN. The equation discovery was also performed for a Bouc-Wen model for a nonlinear hysteretic system. KAN suggested modulus functions that were critical to capturing the system's hysteretic behavior, compared to the conventional choice of functions that can capture the displacement response but not the hysteresis. Much still needs to be explored for the equation discovery of hysteretic systems as they contain an unobservable variable that is critical for capturing the system behavior. Future works need to be focused on obtaining that unobservable behavior to include in the KAN structure for a robust equation discovery in practical scenarios.

\bibliographystyle{unsrt}  
\bibliography{references}

\end{document}